\documentstyle[12pt,aaspp4]{article}

\def\lsim{\lower.5ex\hbox{$\; \buildrel < \over \sim \;$}}
\def\gsim{\lower.5ex\hbox{$\; \buildrel > \over \sim \;$}}

\begin{document}

\title{THE DEEP X-RAY RADIO BLAZAR SURVEY (DXRBS) \\ I. Methods and First 
Results
{\footnote{Based on observations collected at the European Southern 
Observatory, La Silla, Chile; Kitt Peak National Observatory, Cerro 
Tololo Interamerican Observatory, and the Australia Telescope National
Facility}}}

\author{Eric S. Perlman{\footnote{Visiting astronomer at Kitt Peak National
Observatory}}$^{,}${\footnote{Visiting astronomer at Cerro Tololo Interamerican 
Observatory}}}

\affil{Space Telescope Science Institute, 3700 San Martin Drive, 
Baltimore, MD  21218, USA. \\  Email:  perlman@stsci.edu}

\author{Paolo Padovani{\footnote{Visiting astronomer at the European Southern
Observatory, La Silla, Chile}}$^{,}${\footnote{Currently on leave  at the Space Telescope 
Science Institute, 3700 San Martin Drive, Baltimore, MD  21218, USA}}}

\affil{Dipartimento di Fisica, II Universit\`a di Roma ``Tor Vergata'', \\
Via della Ricerca Scientifica 1, I-00133 Roma, Italy.}

\author{Paolo Giommi$^4$}

\affil{SAX Science Data Center, ASI, Viale Regina Margherita 202, 
I-00198, Italy.}

\author{Rita Sambruna$^{2,}${\footnote {Presently at:  Pennsylvania State 
University, Department of Astronomy, 525 Davey Lab, University Park, PA  
16803}}}

\affil{Laboratory for High Energy Astrophysics, Mail Code 660.2 \\
Goddard Space Flight Center, Greenbelt, MD 20771, USA}

\author{Laurence R. Jones}

\affil{School of Physics \& Astronomy, University of Birmingham,
Birmingham B15 2TT, UK.}

\author{Anastasios Tzioumis and John Reynolds}

\affil{Australia Telescope National Facility, CSIRO, PO Box 76, Epping NSW
2121, Australia}


\begin{abstract}

We have undertaken a survey of archived, pointed ROSAT PSPC data for
blazars by correlating the ROSAT WGACAT database with several publicly 
available
radio catalogs, restricting our candidate list to serendipitous flat
radio spectrum sources 
($\alpha_{\rm r} \leq 0.70$, where $S_\nu \propto \nu^{-\alpha}$). 
Here we discuss our survey methods, identification procedure and first
results. Our survey is found to be $\sim 95\%$ efficient at finding
flat-spectrum radio-loud quasars (FSRQs, 59 of our first 85 IDs) and 
BL Lacertae objects (22 of our first 85 IDs), a figure which is 
comparable
to or greater than that achieved by other radio and X-ray survey techniques.

The identifications
presented here show that all previous samples of blazars (even
when taken together) did not representatively survey the blazar
population, missing critical regions of $(L_X,L_R)$ parameter space
within which large fractions of the blazar population lie.
Particularly important is the identification of a large population of FSRQs
($\gsim 25\%$ of DXRBS FSRQs) with ratios of  X-ray to radio
luminosity $\gsim 10^{-6} ~(\alpha_{\rm rx} \lsim 0.78)$. 
In addition, due to our greater sensitivity, 
DXRBS has already more than doubled the
number of FSRQs in complete samples with 5 GHz (radio) luminosities
between $10^{31.5}$ and $10^{33.5} {\rm ~erg ~s^{-1} ~Hz^{-1}}$
and fills in the region of parameter space between X-ray selected
and radio-selected samples of BL Lacs. 
DXRBS is the very first sample to contain statistically significant numbers
of blazars at low luminosities, approaching what should be the lower end of 
the FSRQ luminosity function.

\end{abstract}


\section{Introduction}

Blazars are the most extreme variety of AGN known.  Their signal properties
include irregular, rapid variability; high optical polarization;
core-dominant radio morphology; apparent superluminal motion; flat
($\alpha_{\rm r} < 0.5$) radio spectra; and a broad continuum extending from
the radio through the gamma-rays (these properties are reviewed in
detail by Urry \& Padovani 1995). 
The broadband emission from blazars is dominated by non-thermal processes
(most likely synchrotron and inverse-Compton radiation),
likely emitted by 
a relativistic jet pointed close to our line of sight (as originally
proposed by Blandford \& Rees in 1978).
The beaming hypothesis has been successful in reproducing the
luminosity function of samples of blazars (e.g., Padovani \& Urry 1990,
1991, 1992; Urry, Padovani \& Stickel 1991), allowing the 
derivation of class properties such as the range of Lorentz factors $\Gamma$
and opening angles in the radio band (and also in other bands for BL Lacs). 
However, the small size of these samples (30-50 objects) has prevented 
detailed modeling of the luminosity function, especially at low powers,
yielding considerable uncertainties on the derived parameters. 
 
%

Due to the rarity and low space-density of
blazars, which make up considerably less than 5\% of all AGN (Padovani 1997),  
``pencil-beam'' surveys of the type carried out by, e.g.,
Castander et al. (1996), are poorly suited for finding blazars.  As has
already been noted by Perlman et al.  (1996a), wide-angle surveys with
appropriately restrictive search parameters are much more efficacious.
Indeed, it is thanks to the advent of such surveys that complete
samples of blazars exist today.  
With the advent of modern archival techniques and deeper wide-angle surveys, 
it is now possible to combine survey techniques and 
develop surveys which can sample the blazar luminosity function deeply and
representatively.  Several such projects are currently underway (\S~3).

This paper describes the DXRBS blazar survey and its
goals, and presents our first 85 firm identifications.  The main 
result presented here is that previous blazar samples are not
representative of the blazar class, missing approximately half of the FSRQ
population and a somewhat smaller part of the BL Lac population.  
The newly identified DXRBS blazars expand the range of $L_X/L_R$ values 
found among blazars with emission lines by an order of magnitude, 
and for the first time
samples the low-luminosity end of the luminosity function of FSRQs 
with reasonable statistics.   In Section 2,
we describe the methods used to find blazar candidates and prepare for
optical observations.  Section 3 contains a detailed discussion
on the subject of classification of flat-spectrum radio sources. 
Section 4 describes the results of our optical
spectroscopy and the makeup of the DXRBS blazar sample as of April 1997.   
Section 5 discusses the redshift distribution of 
both the BL Lac and FSRQ subsamples.  Section 6 
contains a discussion of the
properties of the sample, and the implications of these first results
for unified schemes and upon our picture of the blazar class.   
In Section 7, we discuss the topic of
selection effects.  The
conclusions of this work are summarized in Section 8.

\section{Survey Methods}

\subsection{The Catalogs}

The Deep X-ray Radio Blazar Survey (DXRBS) uses
a cross-correlation of all serendipitous sources in the publicly available
database of ROSAT sources, WGACAT (White, Giommi \& Angelini 1995),
having quality flag $\ge 5$ (to avoid problematic detections) with a
number of publicly available radio catalogs. North of the celestial
equator, we used the 20 cm and 6 cm Green Bank survey catalogs NORTH20CM
and GB6 (White \& Becker 1992; Gregory et al. 1996), while south of the
equator, we used the Parkes-MIT-NRAO catalog PMN (Griffith \& Wright
1993). All sources with radio spectral index $\alpha_{\rm r} \leq 0.7$ at a few
GHz were selected as blazar candidates.  

For objects north of the 
celestial equator, 6-20 cm radio spectral indices
were obtained directly from the cross-correlation
of the GB6 and NORTH20CM catalogs. For sources at southern
declinations,  the lack of a comparably deep radio survey at a second
frequency required a different strategy.  In the band $0^\circ >
\delta > -15^\circ$, we cross-correlated the sources with the public
NVSS database (Condon et al.  1997); our selection of
candidates is still not completed in this declination range, since
the NVSS is not yet 100\% complete.  Further
south, the positional accuracy of the NVSS, which covers the sky north of
$-40^\circ$, decreases somewhat (Condon
et al. 1997). In this region, 
we conducted a snapshot survey with the
Australia Telescope Compact Array (ATCA) at 3.6 and 6 cm (note that the 
positional accuracy of the ATCA snapshots deteriorate significantly 
at $\delta >-15^\circ$ 
due to the East-West nature of the array), to get also radio spectral 
indices unaffected by variability (see \S~7). The first set of ATCA
observations (of 163 X-ray/radio sources) took place 11-13 November 1995. 
A second set of 55 X-ray/radio sources (some of which have preliminarily
been classified as blazar candidates based upon the 6-20 cm spectral index
computed from their PMN and NVSS fluxes) were observed in October 1997
to complete the coverage of the southern sample.   These ATCA observations
will be discussed in a future paper.
We had originally requested observations
at 6 and 20 cm at the ATCA as well, to match our northern sample, but the time
allocation committee decided otherwise, based on the instrumental 
configuration. Note that, as the NVSS has a much smaller beam 
size than the GB6 survey, it is preferable to use, whenever possible, 
the NORTH20CM
20 cm fluxes to derive spectral indices. Extra care was taken in the $0^\circ >
\delta > -15^\circ$ region, where we had to resort to the NVSS for this 
purpose, to include the flux from all sources in a 2 arcmin radius. We stress,
however, that this problem is severe only for extended, steep-spectrum 
radio sources, and not for the core-dominated, flat-spectrum sources we 
are interested in. 

\subsection{The Criteria}

We have identified as our highest priority sources those which meet
the following four criteria:

\begin{enumerate}
\item{$\alpha_{\rm 6-20} \leq 0.7$ for $\delta > -15^\circ$; 
$\alpha_{\rm 3.6-6} \leq 0.7$ for $\delta < -15^\circ$}
\item{$|b| >10^\circ$;}
\item{$F(20 {\rm ~cm}) > 100$ mJy for $0<\delta<75^\circ$;}
\item{$F(6 {\rm ~cm}) \gsim 50$ mJy for $\delta<0^\circ$.}
\end{enumerate}

These criteria were chosen in order to ensure that a well defined,
flux-limited sample can be achieved. Over 200 candidate blazars met the 
above defining criteria (details are given below). 
In addition, 98 previously identified but serendipitously observed
objects meet our criteria (see \S~4.3).  Lower-galactic-latitude and
lower-flux sources were assigned to a lower-priority list, but are
useful in order to fill in higher $L_X/L_R$ areas of parameter space.
No pre-selection was imposed on WGACAT in the region $75^\circ \geq \delta
\geq -90^\circ $, although for ease of identification we avoided the regions
within $5^\circ$ of the Large and Small Magellanic Clouds, as well as M31.

Note that although we have not imposed 
any cut in X-ray flux, our sample will have both radio and X-ray flux limits, 
the latter depending on the region of the sky surveyed due to the serendipitous
nature of WGACAT (similarly to the EMSS). Depending on the length of each
individual exposure and the distance from the center of the PSPC field, 
the X-ray flux limits appropriate for each source will vary between 
$\sim 10^{-14}$ and $\sim 10^{-12}$ erg cm$^{-2}$ s$^{-1}$. 
The slightly different radio flux limits north and south of the equator
are simply explained: For $\delta>0^\circ$ (and $\le 75^{\circ}$, the
limit of the GB6 catalog), it is easy to see that all sources with $F
(20 {\rm ~cm}) > 100 $ mJy (the limit of the NORTH20CM catalog) and
$\alpha_{\rm r} \leq 0.7$ will be above the much lower flux limit of the GB6
catalog ($\sim 25$ mJy; note that the converse does not hold). 
For $\delta<0^\circ$, on the other hand, our radio flux limit is the 
completeness 
limit of the PMN survey, which (while declination dependent; see 
Griffith \& Wright 1993) averages about 50 mJy.

\subsection{The Cross-correlations}

The actual cross-correlations were done as follows: north of $\delta >
0^\circ$ WGACAT was correlated with the GB6 catalog with a correlation radius
of one arc minute. The resulting sample, which included 1,119 sources, was
then correlated with the NORTH20CM, this time with a correlation radius of 3
arc minutes, as the positional uncertainties of the NORTH20CM catalog are
considerably worse than those of the GB6 catalog (160 arcsec at the 90\%
level compared to 10-15 arcsec at the $1 \sigma$ level). This
produced a list of 570 sources, 262 of which are unclassified (see below for
details on the classification of WGACAT sources). The 6-20 cm spectral index
was then calculated and 89 sources turned out to have $\alpha_{\rm r} \leq
0.7$ and $|b| > 10^\circ$.
The total number of candidates in the south is still growing. The
correlation of WGACAT with the PMN catalog, with a correlation
radius of one arcmin, produced a list of 541 objects, 310 of which
are unclassified. Of these, 223 are at $|b| > 10^\circ$ and $\delta
< 0^\circ$ (the PMN sample reaches $\delta \sim 10^\circ$). So far, 103
have turned out to have $\alpha_{\rm r} \leq 0.7$, but this number is
bound to increase with the analysis of our October 1997 ATCA observations
and the completion of the NVSS (almost complete as of October 1997).

To evaluate our completeness it is important to derive WGACAT positional 
error circles. This was done as follows.
WGACAT was cross-correlated with the Hipparcos Input Catalog (see e.g., 
Torra \& Turon 1985) and the offsets between the two databases were obtained 
for 6 bins of
the distance from the WGACAT field center (0 - 10 arcmin, 10 - 20 arcmin
etc.).  One $\sigma$ positional errors were then estimated by sorting the
offsets in ascending order and taking the value which included 68\% of the
objects in the bin. These values, reported in Table 3, range from 13 arcsec
for the inner 10 arcmin of the PSPC field to 53 arcsec for the $50 - 60$ 
arcmin ring.

Since the positional accuracy of radio catalogs decreases with flux, we 
investigated the possibility that a one arcminute cross-correlation radius 
might not be large enough at lower radio fluxes and/or large PSPC center 
offsets.
We therefore cross-correlated WGACAT with the GB6 and PMN radio catalogs with a
1.5 arcmin radius. Total positional errors were derived by summing
in quadrature the X-ray and radio uncertainties, the latter obtained from 
the GB6 catalog and from the PMN radio fluxes via the formulae given in the 
PMN papers. The significance of the match was quantified by the 
ratio between X-ray/radio offset and combined positional error. (Note that
1.5 arcmin is roughly equal to twice the combined X-ray and radio uncertainty
of a source with PSPC offset $\sim 30$ arcmin and radio flux $\sim 50$ mJy.)

The results are as follows: for the WGACAT/PMN correlation, the number of
X-ray/radio matches using a correlation radius of 1.5 arcmin increases by
40\%. Dividing the WGACAT sample in an inner (PSPC offset $\le 30$ arcmin) and
outer (PSPC offset $> 30$ arcmin) region, there is a 38\% increase in the
inner region and a 45\% increase in the outer region. In the inner region most
of the increase is due to ``spurious'' associations, which we define for the
purpose of this experiment as those matches with ratio between offset and
positional error $> 2$. Of the 144 new sources, 104 are spurious,
with a net increase of ``good'' sources $\sim 12\%$. In the outer region, of
the 81 new sources, only 13 are spurious, so the number of good sources
increases by $\sim 38\%$. This simply reflects the fact that WGACAT sources
with larger offsets have larger positional uncertainties and ``real'' matches
can have X-ray/radio offsets larger than one arcmin. For the WGACAT/GB6
correlation, the results were slightly different: the increase is only 25\%,
practically independent of WGACAT offset. The net increase of ``good'' matches
is only $\sim 4\%$ in the inner region and $\sim 21\%$ in the outer one. That
is, as the GB6 positions are better than the PMN ones, increasing the
correlation radius has a bigger effect on the WGACAT/PMN ``real'' matches than
it has on the WGACAT/GB6 matches.

As a result of this experiment, we then added to the WGACAT/PMN candidate list
the 29 unclassified sources obtained from the 1.5 arcmin correlation and
having PSPC offsets $\le 30$ arcmin, ratio between X-ray/radio offset and
positional error $\le 2$, $|b| > 10^\circ$ and $\delta < 0^\circ$. 
(Note that these WGACAT/PMN additional candidates all have relatively small 
radio fluxes within a factor 2 of the PMN completeness limit.) By
comparison, only 5 of the additional GB6 sources with PSPC offset $\le 30$ 
arcmin derived from the correlation
with the larger radius had entries in the NORTH20CM catalog but all of them had
$\alpha_{\rm r} > 0.7$. This is easily explained: all these sources had
relatively large positional errors and therefore relatively small radio fluxes
(F(6 cm) $ \sim 20 -30$ mJy). Therefore, only steep-spectrum radio sources
could have a 20 cm flux $> 100$ mJy, the limit of the NORTH20CM catalog. 

In summary, based on the positional accuracy of both WGACAT and the radio
catalogs used to construct our candidate list, we expect our final sample
to be complete in terms of radio/X-ray correlations 
to the radio flux limit for center offsets $\lsim 30$ arcmin. 
At present, these make up about 60\% of our sources.  At larger offsets, our
completeness limit will be at somewhat higher radio fluxes.

We have already mentioned the possibility that some of the X-ray/radio
associations are simply chance coincidences. From the number counts of
flat-spectrum radio sources (Condon 1984), we estimate that only about 20 
$\pm$ 4
objects in our sample are spurious X-ray/radio sources. Most of these will be
singled out because of their large value of the ratio between X-ray/radio
offset and positional error. 

\section{Source Classification}

The definition of the blazar class has varied since the 1978 Pittsburgh
conference, where the terminology was first suggested.  The 
original definition of the class (cf. Angel \& Stockman 1980) emphasized
the dominance of a highly polarized, variable, nonthermal continuum over
other properties.  But in the last twenty years, the definition 
of the class as a whole, as well as various subclasses, has varied,
partly as a result of observational selection.  
A variety of names (e.g., HPQ or highly polarized 
quasar; OVV or optically violent variable) have been applied to some objects, 
usually based upon
finding extreme values of one or more of the signal properties of
the blazar class (\S~1).  A more commonly used set of subclasses are 
based upon the character of the optical 
spectrum:  FSRQ (flat-spectrum radio quasar,
for objects with emission-line dominated spectra)
and BL Lac (nearly lineless objects).  Other authors have restricted
the term ``blazar'' to those with emission-line spectra.  

Further sub-divisions have been invented
to describe objects found as a result of X-ray or radio surveys, or with 
certain broadband spectral shapes (e.g., X-ray selected BL Lacs or XBL, radio 
selected BL Lacs or RBL, low-energy peaked BL Lacs or LBL, high-energy peaked
BL Lacs or HBL; see Padovani \& Giommi 1995a and Urry \& Padovani 1995).  
While the latter two are at least based upon a strictly defined spectral shape
(see \S 6.2), 
all point out the difficulties
inherent in defining the properties of a class based upon 
single-band surveys which cover fairly small ranges of flux in their survey 
band.  Since the blazar population spans
over seven decades of luminosity in the radio, optical and X-ray band, and
over four decades in its ratio of X-ray to radio luminosity, 
single-band surveys are unable
to representatively sample the blazar population, particularly 
when the dynamic range of fluxes being surveyed (i.e. $F_{max} / F_{lim}$) is
less than 100.  

The result is a confusing array of 
nicknames which are utilized with abandon in today's literature.  
The physical meaning of many of these divisions is not at all clear.
For example, much has been
written about the temporary appearance of broad H$\alpha$ with  
$W_\lambda \approx 6$ \AA, in the
spectrum of BL Lac (Vermeulen et al. 1995).  Similar occurrences have
been noted in other objects, including 0846+513 (Arp et al. 1979),
0537$-$441 (Peterson et al. 1976), and 0215+015 (Boisse \& Bergeron 1988).  
Indeed, the recent results of
Sambruna et al. (1996) and Scarpa \& Falomo (1997) suggest that the
separation between BL Lac and FSRQ may be rather ill defined and
perhaps of questionable physical meaning.  And while it is true that 
the properties of BL Lacs found in radio surveys (which mostly have values of
$\alpha_{{\rm rx}} \gsim 0.8$) are considerably different from those found
in X-ray surveys (which mostly have lower values of $\alpha_{{\rm rx}}$),
the explanation for this difference is controversial, and has been
the subject of some debate in the literature (e.g., 
Padovani \& Giommi 1995a; Fossati et al. 1997; 
Georganopoulos \& Marscher 1997).

A variety of deeper, multiwavelength surveys for blazars (of which DXRBS is
one) are currently underway.  The X-ray based surveys, DXRBS (this paper), REX
(Wolter et al. 1997; Maccacaro et al., in preparation), HQS/RASS (Nass et
al. 1996), RC (Kock et al. 1996), RGB (Laurent-Muehleisen et al. 1997), 
and RASS/NVSS (Giommi, Menna \& Padovani, in
preparation) take as their starting point either the pointed ROSAT database
(REX, DXRBS) or the all-sky survey (HQS/RASS, RGB, RC, RASS/NVSS), and make up
their candidate list via cross-correlations with radio survey lists or other
properties. The radio-based surveys
emerging from the FIRST project take the radio-selected FIRST sample as their
starting point, and use variability, polarization and optical colors to 
select candidates (Gregg et al. 1996; Laurent-Muehleisen et al. in prep).
These projects are complementary, using different techniques to sample
different regions of parameter space.  These surveys will both sample the 
parameter space available to blazars more deeply and fill the holes left by 
previous, disjoint selection techniques.
As they do so, we will gain the first complete
picture of the range of properties encompassed by the blazar class.  Given
the confusing array of names currently in use (which may or may not be 
physically meaningful), these surveys (once completed) will need to 
resystematize the blazar definition, as well as those of its subclasses.
However, neither this survey
nor any of the other new X-ray or radio-based surveys can yet undertake the
task of resystematizing the classification of flat-spectrum radio sources, 
as the identification of their samples are all incomplete.  To do so at
this time would risk not only confusion, but the real possibility of missing
a population still extant within the unidentified objects.

For the present paper, we will adopt a form of the FSRQ-BL Lac dichotomy,
basing our classifications solely upon the optical spectrum.  We will apply
the term ``blazars'' to both BL Lacs and FSRQs, since recent evidence (Fugmann
1988; Impey, Lawrence \& Tapia 1991; K\"uhr \& Schmidt 1990; Jannuzi et
al. 1993, 1994) has shown that the properties outlined in \S~1 are shared by
both FSRQs and BL Lacs. We adopt the modified form of the BL Lac definition
advocated by March\~a et al. (1996; see their Fig. 6) to classify BL Lacs and
radio galaxies.

Their starting point is that the line luminosity seems to be independent of the
observed continuum in blazars (see Koratkar et al. 1998 for an example of this
behavior in 3C 279). It then follows that the Ca H \& K break contrast, a
measure of the presence of non-thermal continuum in a galaxy [defined by $C =
(f_+ - f_-) / f_+$, where $f_+$ and $f_-$ are, respectively, the flux redward
and blueward of the Ca break], and equivalent width $W_\lambda$ will be
correlated (the lower $C$, i.e., the higher the non-thermal contribution, 
the lower $W_\lambda$). Thus, changing the viewing
angle and/or the luminosity of the BL Lac relative to that of the host galaxy
will move an object on a diagonal trajectory in the contrast -- equivalent
width plane. March\~a et al. (1996) showed convincingly that objects in a
triangular area limited by contrast $C = 0.4$ (breaks with $C \sim 0.5$ are
typical of elliptical galaxies; Dressler \& Schectman 1987) and the diagonal
line shown in Figure 1 (which assumed the line and galaxian continuum emission
of 3C 371 as its starting points and a smoothly decreasing AGN contribution)
should still be called BL Lacs. Note that this expands
upon the ``classical'' definition used by previous authors (Stickel et al.
1991; Stocke et al. 1991; Perlman et al. 1996a) of equivalent width $W_\lambda
< 5 (1+z)$ \AA~ for all emission lines and Ca H \& K break strength $C <
0.25$. The fuzziness of the previously used criterion is further illustrated
by the occasional observation of broad H$\alpha$ lines in the spectra of
several famous BL Lacs (among them Mkn 501 and BL Lac; see, for example
Vermeulen et al. 1995). Objects which fall outside the traditional definition
of the BL Lac class, but within the March\~a et al. definition which we adopt,
will be discussed individually in \S~4, and we will return to the subject in
\S~7 when selection effects are discussed.


Thus, objects which meet the March\~a et al. criteria are classified herein
as BL Lac objects.  Objects with higher-equivalent-width emission lines which
are still narrow (FWHM $\lsim 1000-2000 $ km/s), or stronger Ca H \& K breaks,
are classified as radio galaxies, and are discussed individually in \S~4, and
as a group in \S~7.
All flat radio spectrum objects with higher equivalent widths and broad
emission lines (FWHM $\gsim 1000-2000 $ km/s)
are classified as FSRQs.   It is important to note that this classification is
being applied without regard to any other characteristic, such as redshift, 
presence or lack thereof of a stellar continuum, or physical extent.  
As a result, a number of objects which have been called broad-line radio 
galaxies by other authors are included as FSRQs within 
the previously-identified 
portion of our sample.  We return to this last topic in \S~7.

Due to the many commonalities shared by all blazars, we believe that a
unified approach to these enigmatic objects is more helpful in helping us
understand them, similar to that taken for radio galaxies by Baum,
Zirbel \& O'Dea (1995) and Zirbel \& Baum (1995).  
Therefore, while we will use the classical definition to classify sources
here (in order to ensure easy compatibility with past studies), 
it is our goal in future works to consider the 
equivalent width and luminosity of emission lines simply as additional
variables in the analysis.  

\section{Sample Identification}

In order to identify a candidate object as either a BL Lac, FSRQ or radio
galaxy, an optical spectrum is required.  
To pinpoint the optical counterpart we used, where available, positions from
either the NVSS or our ATCA survey (both of which have errors $\lsim 3''$ 
relative
to the Digitized Sky Survey; note that the figure we use 
is significantly larger 
than typically quoted for NVSS and ATCA detections with $F>50$ mJy due to the 
non-planarity in the sky survey plates and the slight inconsistencies 
between the coordinate systems of each; see, e.g., Irwin, Maddox \& McMahon 
1994;  Drinkwater et al. 1995) to
obtain finders from the Digitized Sky Survey (using Skyview, McGlynn \&
Scollick 1996).  For some sources which did not have NVSS positions
before the time of optical observations, arcsecond positions from the Texas 
survey (Douglas et al. 1996) and NED (typically based on VLA data) were used.  
Magnitudes for all X-ray/radio sources with counterparts 
on the POSS and UKST plates
which comprise the Digitized Sky Survey were obtained from the
Cambridge APM and Edinburgh 
COSMOS projects (Irwin et al. 1994; Drinkwater et al. 1995), 
except where a blend of two
or more sources were observed, in which case a magnitude was estimated
by eye.  All X-ray/radio sources without counterparts on the survey
plates were imaged at either the KPNO 0.9m or the CTIO 0.9m
telescopes.  This allowed identification of all optical counterparts to
R = 23.

Spectroscopic observations  were conducted at
the KPNO 2.1 m, MMT, Lick 3 m, ESO 2.2 m and 3.6 m, and CTIO 1.5 m
telescopes.  One object, WGAJ0449.4$-$4349, was observed by M. Ruiz at the
CTIO 4 m telescope in January 1996; Dr. Ruiz has kindly allowed us to publish
these data herein.  In Table 1, we list all telescope runs (including, for
completeness, the MMT and Lick runs, which were obtained for other
projects,  but during which a few DXRBS blazars were observed) and
relevant details  of the observing setup, such as approximate 
wavelength range and
resolution.  In Table  2, we list the details of the observations. 
Due to poor weather during the
KPNO run (2 mostly cloudy nights out of 3) the data from this run are
of lower quality than the ESO data.  With the exception of the CTIO 4m and 
MMT observations, spectra were generally not taken at parallactic angle (mostly
due to the difficulty of changing the position angle of the slit at the 
KPNO 2.1m, CTIO 1.5m and ESO 2.2m telescopes).

The spectra were reduced using standard IRAF routines.  Data were
overscan and bias-subtracted, and flatfielded using programs in the
IRAF package {\it noao.imred.ccdred}, and spectra were extracted,
wavelength-calibrated and flux-calibrated using programs in the package
{\it noao.twodspec}.  Cosmic rays were removed in the 1 and 2-dimensional
data by hand.

A dereddening correction was applied to the data
using the IRAF routine {\it noao.onedspec.dered} and assuming Galactic values
of extinction derived from 21-cm measurements (Stark et al. 1992; Shafer et 
al., private communication). 

We recorded the central wavelength, 
equivalent width, full-width at half-maximum, and flux in each spectral line.
Those data will be given and analyzed in future papers.  
Except where
noted, where only a single emission line was observed, it was assumed to 
be Mg II $\lambda 2798$ \AA. Seven of 22 newly identified BL Lacs
lack recognizable spectral features.  These objects are not included in the 
redshift distributions discussed in \S~5, and we have not computed luminosities
for them. 
Higher signal-to-noise spectroscopic observations are required to obtain 
redshifts for these objects.

\subsection{Identifications and Efficiency}

In Figure 2, we show the spectra of the optical counterparts to 
85 of 97 observed X-ray/radio sources.
All spectra have been smoothed with Gaussians of width 3 pixels.
Twelve spectra are not shown, either because (1) too few photons were
observed to allow a reliable classification to be made (6 sources), (2)
the optical survey plate did not contain enough information to
distinguish whether the counterpart was a bright star near the radio
position or a fainter extragalactic object at the radio position (5
objects), or (3) because a lower-quality radio position was used (1
object).  In this last case (WGAJ1022.1+4126), the object is also very
close to a third-magnitude star, SAO 43310, making spectroscopic
observations difficult. 

Positional information for all 85 sources
for which we announce identifications herein are given in Table 3, including
information from WGACAT, the PMN and Green Bank surveys, the NVSS and our ATCA
survey.  A number of sources were serendipitously observed by ROSAT on more 
than one occasion; for completeness, we give WGACAT positions for all 
observations of DXRBS sources.

Of the 85 newly identified sources, 59 are FSRQs, and 22
are BL Lacs.  Hence, our technique is $\sim 95 \%$ efficient at
selecting blazars, where we define the efficiency as the fraction of objects
which turn out to be blazars in a given survey {\it after} selection criteria
have been applied.  Three of 85 objects are radio galaxies, with CaII
breaks stronger than typical BL Lacs (but see \S~7.1); 
while one quasar, which was 
observed before we had information on its spectral index, turned out to have 
$\alpha_{\rm r} > 0.7$. Classifications, redshifts and observational details
for these sources are given in Table 4.  The 0.1-2.0 keV X-ray fluxes
given in Table 4 are not corrected for Galactic absorption; however, the 
1 keV X-ray fluxes given therein are unabsorbed.  Note that both the 0.1-2.0
keV and 1 keV X-ray fluxes have been derived from ROSAT count rates using the 
observed hardness ratio and assuming Galactic $N_{\rm H}$. These numbers may 
change
somewhat when a more thorough analysis of the X-ray spectrum is done (this is
in progress).  For objects
observed more than once by ROSAT, we give in Table 4 the count rates and 
X-ray fluxes found for each observation. For objects which we classify as 
either radio galaxies or BL Lacs, we give the equivalent width of the 
strongest emission line and Ca break strength in Table 5; these
values have been also been displayed in Figure 1. 


Previous X-ray and radio surveys can claim efficiencies which are
nearly comparable to ours.  
For example, the efficiency of the Slew Survey at identifying HBLs within
the well-known $(\alpha_{\rm ox},\alpha_{\rm ro})$ box 
(Perlman et al. 1996a) is
80$-$90\%, and the fraction of BL Lacs and FSRQs within the flat-spectrum
subset of the 1 Jy survey is nearly 90\% (260/298) for a dividing line at 
$\alpha_{\rm r} = 0.5$, but goes below 80\% (284/364) for 
a dividing line at $\alpha_{\rm r} = 0.7$ (Stickel et al. 1994;
see also \S~7).  However, each of these survey techniques were insensitive
to large portions of the blazar population (\S~3).   
The combination of high sensitivity in both the X-ray and radio bands plus 
a two-band survey method gives DXRBS significant advantages over these
previous survey methods. 

\subsection{Comments on Individual Sources}

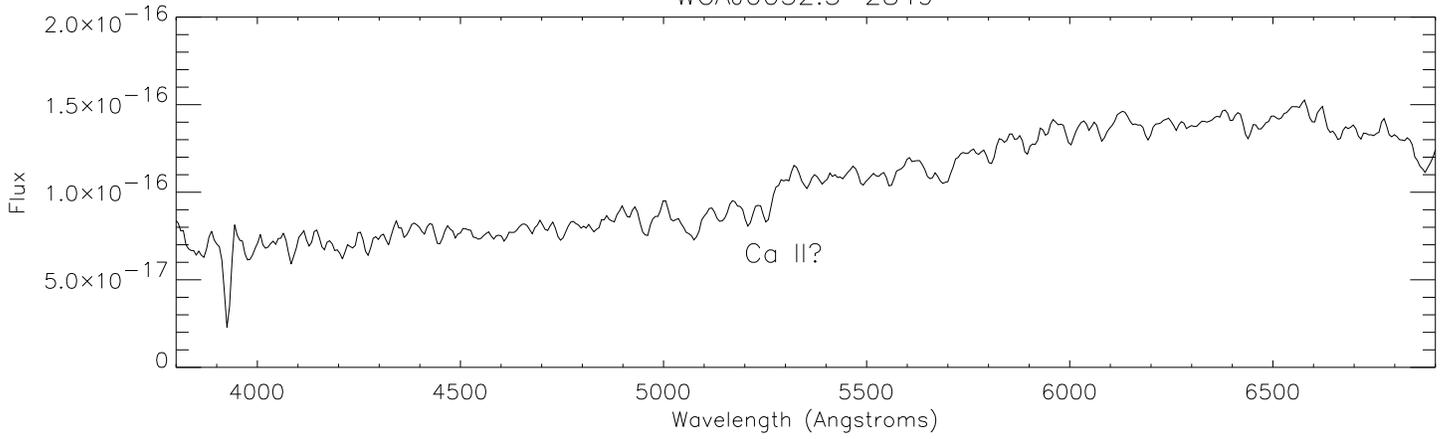
{\bf WGAJ0043.3$-$2638.}  Since being selected for the DXRBS sample, 
this source was observed by both Cristiani et al. (1995) and Wolter et al.
(1998), both of whom identify it as a broad-emission line AGN at 
$z=1.002$.  Both this redshift and the redshift we list in Table 3 ($z=0.451$)
have problems reproducing some of the features found in both spectra.  For 
example, if the object is at $z=1.002$, it is difficult to explain the
likely emission line at 4050 \AA, which we have classified as Mg II at 
$z=0.451$.
Further observations are needed to determine the correct redshift of this
object.  The rest-frame 
equivalent width of the 4050 \AA~emission line is 9.6 \AA~(if $z=0.451$) or
7.0 \AA~(if $z=1.002$), only slightly above the dividing line between BL Lac
and FSRQ.  Therefore, even though the line is clearly broad (FWHM=3800
km/s), this object
is similar to objects such as Mkn 501 and BL Lac which share this property
and must be classified as a BL Lac.

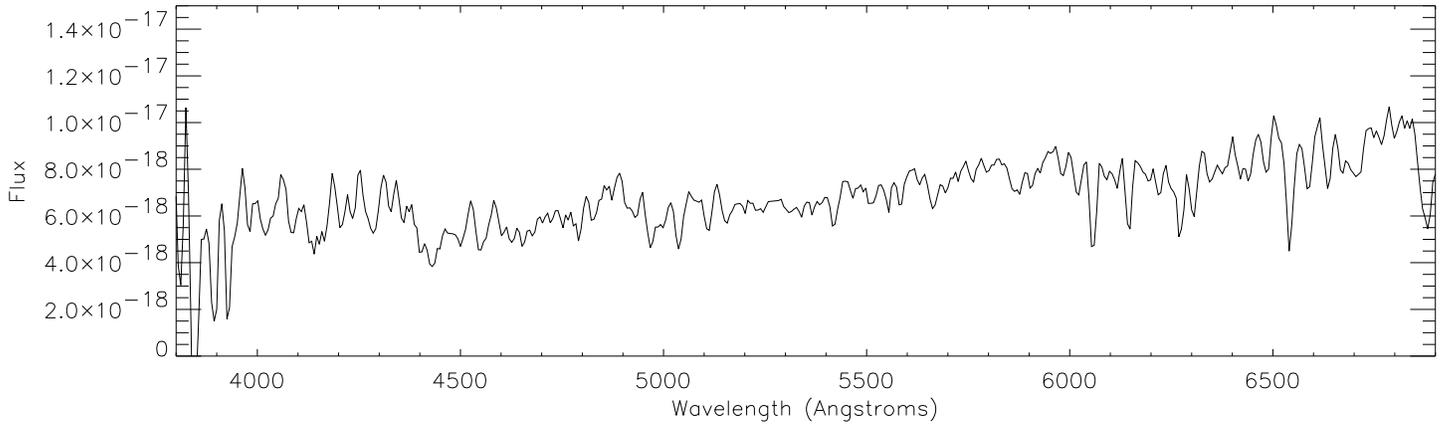
{\bf WGAJ0100.1$-$3337.} The single emission line, which we classify as Mg II
at $z=0.875$, is clearly broad (FWHM=2900 km/s) but its equivalent width is
close to the BL Lac/FSRQ dividing line ($W_\lambda = 9.8$ \AA~rest-frame). 
We classify this object as a BL Lac, similarly to WGAJ0043.3$-$2638 (see 
above discussion).

{\bf WGAJ0204.8+1514.}  This
source, also known as 4C +15.05, has a radio flux $> 3$ Jy, and was also
previously observed by Stickel et al. (1996), who classified the source
as an AGN at $z=0.833$ based upon the identification of two lines as
OII $\lambda$3727 and Ne I $\lambda$3833.  These lines are also
present in our spectrum, as are four other lines (Figure 2).  However,
the redshift claimed by Stickel et al. (1996) is likely incorrect, as all six
lines cannot be accounted for if the redshift is $z=0.833$.
We believe that a better fit is obtained with a redshift $z=0.405$.
This object is also the
likely counterpart of the EGRET source 2EG0204+1514 (Thompson et al.
1995; Mattox et al. 1997).

{\bf WGAJ0210.0$-$1004.} This object, at $z = 1.976$, is $\sim 2'$ from 
MS0207.4$-$1016, identified
by Stocke et al. (1991) as a radio-quiet QSO (F(6 cm) $<$ 0.3 mJy at 3 
$\sigma$) at $z=1.970$.  
A 6cm VLA survey done during the EMSS project showed that there are two fairly
strong radio sources which likely would be in the PMN beam (Stocke, private 
communication). The stronger source, with a flux of 133 mJy, is at the 
position of WGAJ0210.0$-$1004; however, there is another 70 mJy source at a
position which is not consistent with either WGAJ0210.0$-$1004 
or MS0207.4$-$1016.
Even with this reduced 6 cm flux, WGAJ0210.0$-$1004 is still a
flat spectrum source ($\alpha_{\rm r} = 0.58$). There is no 
question about the correctness of either X-ray source identification, 
since WGACAT lists 
a 0.01 ct/s X-ray source at a position consistent with MS0207.4$-$1016 (in 
addition to WGAJ0210.0$-$1004).  It is possible, however, that X-ray emission
from both sources may have contributed to the EMSS X-ray flux. What is
particularly interesting is that these two objects are at essentially
the same redshift, and are therefore likely associated with one another in a 
group or
cluster of galaxies, since the projected separation between them is 1.6 Mpc.

{\bf WGAJ0245.2+1047.}  This object is difficult to classify because of
the large equivalent width of its H$\alpha$ emission line (19.1 \AA~
rest-frame).  However, when combined with its low Ca H \& K break strength 
($C = 0.26$) it rests securely in the BL Lac area of the $(W_\lambda,C)$
plane as defined by March\~a et al. (1996).  We therefore classify this object
as a BL Lac.

{\bf WGAJ0313.9+4115.}  The H$\alpha$ emission line in this object's spectrum
is not very broad, exhibiting FWHM = 1780 km/s.  The rest-frame
equivalent width of this
line is close to the BL Lac/FSRQ dividing line (13.0 \AA~rest-frame).  
The Ca II break strength is 0.38, close to the BL Lac/radio galaxy border
we are using.  We have classified this object as a BL Lac object, but we note
that the $1\sigma$ errors on our measurement of $C$ are not small enough to
exclude the alternate classification as a radio galaxy (Figure 1).  A higher 
signal-to-noise spectrum of this object is clearly necessary to confirm 
its nature.

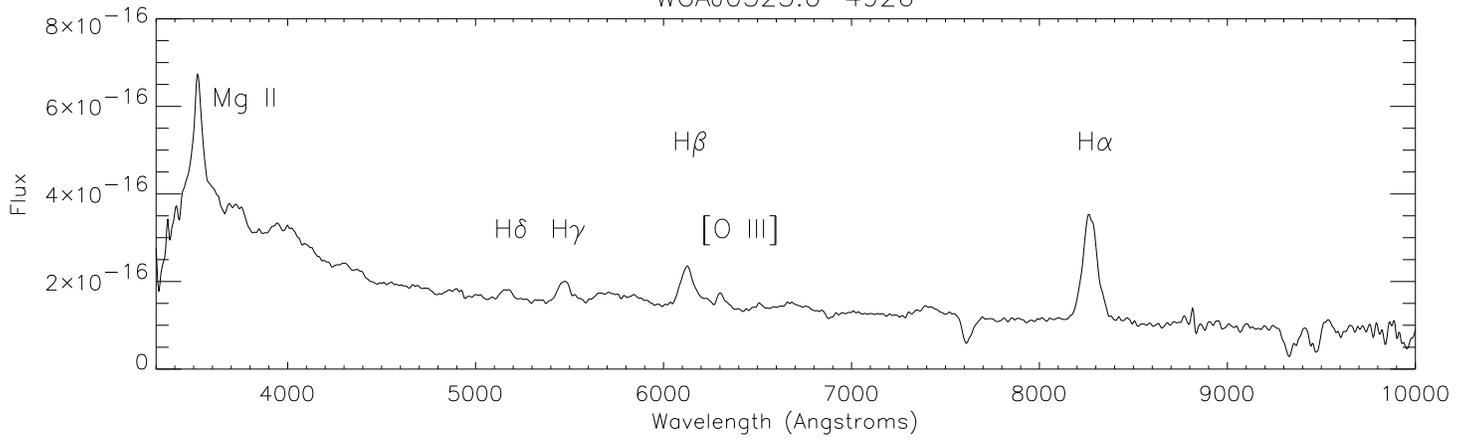
{\bf WGAJ0340.8$-$1814.} The H$\alpha$ emission line in the spectrum of this
object has a rest-frame equivalent width of 16.0 \AA, and its Ca II break
strength is $C=0.40$.  Thus it is right on the borderline of the BL Lac
region of the $(W_\lambda,C)$ plane.  We have classified this object as a
radio galaxy; however, a higher signal-to-noise spectrum is clearly necessary
to confirm its nature.

{\bf WGAJ0421.5+1433.}  Our spectra show no clear lines; however, due to its
low signal-to-noise ($\sim 7$), the 2$\sigma$ upper limits that can be placed
on its break strength ($C<0.30$) and equivalent width of emission lines 
($W_\lambda < 8.2$ \AA) are not very stringent.  They are adequate, however, to
allow us to classify this object as a BL Lac.  Better observations of this 
source are clearly necessary.

{\bf WGAJ0428.8-3805.}  This object has no detectable emission lines in its 
spectrum (2 $\sigma$ upper limit = 0.7 \AA), and a weak Ca break ($C = 0.32$).
We classify it as a BL Lac object using the March\~a et al. (1996) criteria.

{\bf WGAJ0449.4$-$4349.} This bright BL Lac object was observed as a target by
ROSAT; however, until now it was unidentified. Due to its nonserendipitous 
observation by ROSAT, we will not include it in computations of the luminosity
function.  We include it here as it was identified during our observing
campaign and there will probably be no other opportunity to discuss it.

{\bf WGAJ0500.0$-$3040.}  We have termed this object a radio galaxy despite the
fact that all of its emission lines have equivalent widths greater than 
5 \AA~(some are as large as 70 \AA), since all are relatively narrow (FWHM 
$\sim 1000-2000$ km/s). However, there is no detectable 4000 \AA~break in its 
spectrum, which points to an unusually strong non-thermal contribution. 

{\bf WGAJ0513.8+0156.}  The Ca II break strength ($C = 0.34$) and lack of 
emission lines (2 $\sigma$ upper limit on $W_\lambda = 1.3$ \AA) allow us to
classify this object as a BL Lac object.  However, the low signal-to-noise of
its spectrum blueward of the Ca II break results in a relatively large 
1 $\sigma$ error on its break strength, large enough so that we cannot exclude
the alternate classification as a radio galaxy (Figure 1).  A higher 
signal-to-noise spectrum of this source is required to confirm its nature.

{\bf WGAJ0558.1+5328}.  The H$\alpha$ emission line in this object's spectrum
is not very broad, exhibiting FWHM = 2100 km/s.  The equivalent width of this
line is close to the BL Lac/FSRQ dividing line (9.8 \AA).  The Ca II break
strength is 0.29.  We classify this object as a 
BL Lac.  However, the low 
signal-to-noise of its spectrum, particularly blueward of the Ca II break
($\sim 5$ compared to $\sim 12-15$ redward of the break) do not quite allow
us to exclude the alternate (radio galaxy) classification.  A higher 
signal-to-noise spectrum would clarify this question.

{\bf WGAJ0624.7$-$3230.}  The emission line and absorption line redshifts of
this object are somewhat different ($z_{abs} = 0.252$ and $z_{em} = 0.275$).
Therefore it is likely that the galactic emission in the spectrum is due to
a foreground galaxy superposed upon the radio source.  The sole emission line
has a rest-frame 
equivalent width of 8.5 \AA; slightly over the BL Lac/FSRQ dividing 
line, but the line is narrow (FWHM = 900 km/s).  We are classifying this 
object as a BL Lac because of its small equivalent-width emission line 
and low Ca H \& K 
break contrast ($C = 0.22$).

{\bf WGAJ0656.3$-$2403.}  There is a possible emission feature in the spectrum
of this object at 3845 \AA.  It is unclear whether this is a real emission
line, noise, or a cosmic ray due to the noisiness of the spectrum in this 
range.  The lack of other emission lines in the spectrum and high noise level
in the blue make it somewhat
doubtful that this feature is truly an emission line.  However, if it is due
to Mg II emission, the redshift of this object would be $z=0.371$, and it 
would be narrow (FWHM = 500 km/s).  We classify this object as a BL Lac 
due to the likely lineless nature of its spectrum, though we note that if
the 3845 \AA~feature is indeed an emission line it exceeds by more than a 
factor of four the traditional BL Lac/FSRQ dividing line (rest-frame
$W_\lambda = 24.5$ \AA).

{\bf WGAJ0724.3$-$0715.}  Despite its faintness, the H $\alpha$ emission line 
in the spectrum of this object is quite broad (rest-frame $W_\lambda = 30.3$ 
\AA, FWHM = 4000 km/s).  We therefore classify it as an FSRQ.

{\bf WGAJ0744.8+2920.} This object was identified independently by Gregg et al.
(1996) as part of the FIRST bright QSO sample, and by Wolter et al. (1998).  
We confirm both identifications
and redshifts; however, a comparison of our spectrum with that given by Wolter
et al. reveals a large deficit in the blue in our spectrum.  This is most 
likely due to a combination of instrumental and weather related factors.

{\bf WGAJ0816.0$-$0736.} We tentatively classify this object as a BL Lac due to
its lack of emission lines and low break strength ($C=0.37$).  However, due to
the low signal-to-noise of its spectrum blueward of the Ca II break ($\sim 4$
compared to $\sim 20$ at $>5000$ \AA), we cannot exclude the alternative 
(radio galaxy) classification due to the large 1$\sigma$ error on $C$ (0.18).
A higher signal-to-noise spectrum is required to clarify its nature.

{\bf WGAJ0900.2$-$2817.} A second spectrum of this object, 
with much wider wavelength
coverage, was obtained in May 1997 at the ESO
2.2m. That spectrum (which will be published in a later paper) 
confirms the identification of the single line as 
Mg II $\lambda 2798$.  

{\bf WGAJ1057.7$-$7724.} The fairly low signal-to-noise spectrum we have ($S/N 
\sim 7$) places 2 $\sigma$ limits on $C$ and $W_\lambda$ which are sufficient
to classify this object as a BL Lac. A higher signal-to-noise spectrum is 
required to obtain a redshift.

{\bf WGAJ1222.6+2934.} The fairly low signal-to-noise spectrum we have ($S/N 
\sim 6$) places 2 $\sigma$ limits on $C$ and $W_\lambda$ which are sufficient
to classify this object as a BL Lac. A higher signal-to-noise spectrum is 
required to obtain a redshift.

{\bf WGAJ1525.3+4201.} This source was listed as a BL Lac candidate by
Ruscica et al. (1996). However, our spectrum shows strong, broad lines, and we
identify this object as a quasar at z = 1.189.  

{\bf WGAJ2317.9$-$4213.} This object is most likely a radio galaxy based upon
its strong Ca II break ($C=0.52$).  It is probably associated with a group
of galaxies at the same redshift found by the Las Campanas Redshift Survey
(Schechtman et al. 1996).

{\bf WGAJ2322.0+2114.}  This object was also observed by Wolter et al. (1998).
We confirm both their identification and redshift.

\subsection{Previously Identified Sources}

Previously known serendipitous sources 
were selected by cross-correlating WGACAT with a variety of optical and radio
catalogs, including the Ver\'on-Cetty \& Ver\'on (1996) and Hewitt \& Burbidge
(1993) quasar catalogs, the 1 Jy (Stickel et al. 1994) and S4 (Stickel \&
K\"uhr 1994) radio catalogs, and the BL Lac catalog of Padovani \& Giommi
(1995b). Also, the classification of non-AGN sources (which is 
important to select the unclassified objects) was done as described in 
White et al. (1995). 
In some cases classifications were also double-checked in the NASA
Extragalactic Database (NED) or taken from very recent papers. All objects
broad-line radio galaxies (BLRGs) have been
included with the FSRQ class (see \S 3). The adopted cross-correlation radii 
were the same as those used for the selection of our candidates (\S~2).  
Selection criteria identical to those described in \S 2 were applied to the
previously known objects.  
Potential mis-identifications through chance coincidences of previously known
AGN which fulfill our selection criteria were addressed by shifting the X-ray
positions by one degree at a time several times, and repeating the
cross-correlations between WGACAT and the AGN catalogs. The number of spurious
X-ray/optical associations was zero, which implies a negligible (2 $\sigma$
upper limit $\sim 4\%$) contamination by spurious sources. 


Ninety-seven previously known objects were thus found, of
which 11 are BL Lacs and 76 are FSRQs. The remaining ten objects are
Narrow Line Radio Galaxies (NLRGs) with $\alpha_{\rm r} \le 0.7$. 
Details of these 97 objects are given in Table 6. 
The mean X-ray/optical offset for these sources
is $\simeq 22$ arcsec, while the median one is $\simeq 17$ arcsec, in
agreement with the estimated errors on the positions of the WGA sources (\S~2).
Since no preselection was done on WGACAT for these
objects (with the exception of excluding regions at $|b| < 10^\circ$ and 
within $5^\circ$ of the LMC, SMC and M31), these objects will qualify for
our complete sample, and will be discussed along with the newly-identified
sources in \S \S 5-7.

\subsection{Sample Properties}

In this subsection we utilize the data given in Tables 4 and 6
to calculate basic parameters such as X-ray and radio
luminosities.  The distributions of these parameters will be analyzed
in upcoming sections.  In Figure 3, we plot the 
redshift distribution of DXRBS FSRQs opposite that of the 1 Jy and S4 samples.
In Figure 4, we plot the redshift distribution of DXRBS BL Lacs opposite the
Slew and 1 Jy samples.
In Figure 5, we have plotted the X-ray and radio luminosities
of all the FSRQs in our sample, as well as those of the 1 Jy and S4 samples.
A similar plot is given in Figure 6 for the BL Lacs, with the comparison 
samples being the 1 Jy, Slew and EMSS samples.  The apparent deficit of
objects at the very highest luminosities is a result of the smaller
area of sky covered by DXRBS compared to, e.g., the Slew and 1 Jy samples.
The 1 keV 
X-ray luminosities plotted in Figs. 5 and 6 have been $K$-corrected and
de-absorbed using the X-ray spectral indices derived from the WGACAT 
hardness ratios as detailed in Padovani \& Giommi (1996).  Where an object 
was observed more than once by ROSAT, the X-ray luminosity plotted represents
the average luminosity.

When complete, the DXRBS blazar sample will include more than 300
blazars, considerably larger than any previous complete sample of
blazars.  Combining the objects for which we have announced
identifications in this paper with previously identified objects, the
sample is now over 50\% identified. All further analysis in this paper 
will be done using all identified objects with redshifts
(including both newly and previously identified objects).

\section{Redshift Distribution}

The distribution of redshifts among the DXRBS FSRQs identified so far (Figure
3) is quite similar to that of the 1 Jy and S4 samples. Not much more can be
said at present. A more thorough comparison will have to await completion of
our survey and the convolution of the redshift distribution with the WGACAT
sky coverage. As has
already been mentioned by Hook et al. (1996), selecting flat-spectrum
sources is an efficient way of finding high-redshift, radio-loud
quasars. We would therefore expect that the majority
of the FSRQs would be  at high redshifts ($z > 1$), a suspicion which
the data confirm.   The mean redshift of our FSRQ sample is $z\sim
1.2$, and the high-redshift tail extends to $z \sim 4$. Our redshift
distribution cannot, however, be directly compared to that of Hook et
al. (1996) because those authors required that the sources be
particularly ``red,'' a criterion we do not impose. 

While a significant fraction (7 of 33) of the BL Lacs so far included in our
sample still lack redshifts, it is apparent that the redshift distribution for
the DXRBS BL Lacs (Figure 4) is dominated by objects at $z<0.4$, and therefore
more similar to that of the Slew Survey (Perlman
et al. 1996a; or the EMSS survey: Stocke et al. 1991) than the 1 Jy
sample (Stickel et al. 1991), which is less peaked at low redshifts and has a
much more prominent high-$z$ tail than either the DXRBS 
or Slew Survey samples. 
As above, a direct comparison between these distributions is not
possible at present, given also the different survey methods and selection
criteria. We only mention here that the difference between the DXRBS and 1 Jy
redshift distributions appears to be striking considering that
the radio flux limit of DXRBS is a factor 20 lower so that if anything, one
would expect a {\it higher} fraction of high-$z$ sources, rather than the
dearth of high-$z$ sources that we observe. The flatter distribution of the 1
Jy sample might be related to the fact that it misses some low
redshift objects hidden within bright hosts (which might be misidentified as
radio galaxies; see \S 7) and might also be contaminated by
misidentified quasars at high redshifts (e.g., 1 Jy 1308+326), as suggested by
March\~a \& Browne (1995) and Perlman et al. (1996b). Some 1 Jy BL Lacs might
be gravitationally lensed if superposed by chance onto a lower-redshift galaxy
(as originally suggested by Ostriker \& Vietri 1985).  This last possibility
now appears quite likely from the results of Stocke \& Rector (1997), which
show that the 1 Jy BL Lacs have a $2.5-3 \sigma$ excess of Mg II absorbers
along their sight lines.

\section{A More Complete Picture of the Blazar Class}

DXRBS' unique combination of a dual X-ray/radio survey method and high
sensitivity has opened up important, large new regions of parameter
space for both blazar sub-classes, within which a large fraction of our
sample (nearly 50\%) lie.  
This section is devoted to discussing these discoveries and their 
impact upon our understanding of the blazar class.

\subsection {FSRQs}

An examination of Figure 5 shows that the FSRQs in the DXRBS sample cover a
much wider range of parameter space than those in the two previously existing
complete samples of FSRQs, the 1 Jy and S4 samples
(radio data for these samples were taken from Stickel et al. 1994 and 
Stickel \& K\"uhr 1994, while X-ray data for these objects were taken from
the multifrequency AGN database of Padovani et al. (1997b; and references 
therein); 
it should be noted that 
the deeper S5 sample is $\sim 25\%$ unidentified and half of the objects lack
redshifts, as noted in Stickel \& K\"uhr 1996).  We have quantified these 
differences by 1-dimensional and 2-dimensional Kolmogorov-Smirnov (K-S) tests.
The 2-dimensional K-S test reveals that the differences in the ($L_X,L_R$)
plane coverage between the DXRBS and 1 Jy and S4 samples are significant: the
probability that the 1 Jy and DXRBS samples could emerge from the same parent
population is 0.2\%, where as the probability that the S4 and DXRBS samples
could emerge from the same parent population is 2.2\%. Given the fainter flux
limits of DXRBS, this is expected.

One-dimensional K-S tests reveal that largest difference is in the radio 
luminosity. The probability that the 1 Jy and DXRBS radio luminosity 
distributions could emerge from the same parent population is $<0.1\%$, and
the probability that the S4 and DXRBS radio luminosity distributions could
emerge from the same parent population is 0.7\%. The mean of the DXRBS 
$L_R$ distribution is different from that of both the S4 and 1 Jy at greater
than 99.9\% significance. The situation is somewhat different for the X-ray
luminosity distribution. The probability that the 1 keV luminosity 
distribution of the 1 Jy and DXRBS sample could emerge from the same parent 
population is 15\%, i.e. our results are inconsistent with them emerging
from a different parent population. The result is similar for the S4 (23\%
probability). Also, the mean of the DXRBS sample's X-ray luminosity 
differs from that of both the 1 Jy and S4 samples' only at the  
$93$ and $92\%$ level respectively. Note, however, that X-ray data are 
available only for $\sim 53\%$ and $66\%$ of the S4 and 1 Jy FSRQs
respectively, so their X-ray luminosity distributions are likely to be skewed
towards the most luminous X-ray sources.

Inspection of Figure 5 reveals that the differences lie in two areas:
at low luminosities (particularly low radio luminosities) and high ratios
of  $L_{\rm X}/L_{\rm R}$.  The former regime could not be surveyed
well by previous surveys due to their considerably higher flux limits.
It is therefore not surprising that, as shown in Figure 5, the 1 Jy and S4
samples together have only twelve objects at radio luminosities 
$L_R < 10^{33.5} {\rm ~erg ~s^{-1}}$ ($\sim 3\%$), and none at   
$L_R < 10^{32.5} {\rm ~erg ~s^{-1}}$. The fraction of
low-luminosity objects is much higher in the DXRBS sample (Fig. 5), which, 
while still incomplete, already contains over twice as many objects 
(28; or $\sim 20\%$) with $L_R < 10^{33.5} {\rm ~erg ~s^{-1}}$, six of which
are at $L_R < 10^{32.5} {\rm ~erg ~s^{-1}}$. The DXRBS
sample is therefore the very first sample of blazars to contain statistically
significant numbers of blazars at low luminosities, approaching
what should be the lower end of the FSRQ luminosity function according
to unified schemes, i.e. $\sim 10^{31.5} {\rm ~erg ~s^{-1} ~Hz^{-1}}$
(Urry \& Padovani 1995).

The discovery of a large population of FSRQs with ratios of X-ray to radio
luminosity $L_X/L_R > 10^{-6}$ ($\alpha_{\rm rx} < 0.78$), values more 
similar to HBLs, is more startling, as few such objects were known in 
previous complete samples (there are nine such objects in the 1 Jy and S4
combined; see Fig. 5).  The finding of a large population 
of ``HBL-type'' FSRQs contradicts the prediction of Sambruna et al. (1996)
that, based upon the similarities in the optical-X-ray broadband
spectral characteristics of LBLs and FSRQs, there should be no HBL-type
FSRQs.  Padovani, Giommi \& Fiore (1997b) were the first to
notice that about 17\% of all radio quasars 
with radio/optical/X-ray data (previous to DXRBS) fell in the region of the 
$(\alpha_{\rm ox},\alpha_{\rm ro})$ plane typical of HBLs (or X-ray selected
BL Lacs) and called them ``HBL-like'' quasars. 

We term these objects ``HFSRQs'', or
high-energy peaked FSRQs; this terminology stresses their apparent similarity
to the HBLs. These objects comprise $\sim 25\%~(32/135)$ of the DXRBS sample 
of FSRQs so far. However they probably comprise a somewhat larger proportion 
of the DXRBS FSRQ population as a whole, as $\sim 40\%$ (25/59) of the 
newly-identified FSRQs are HFSRQs.
Padovani et al. (1997b) have proposed that the X-ray band in these objects,
unlike in lower $L_X/L_R$ FSRQs, in which inverse Compton emission
prevails (Padovani, Giommi \& Fiore 1997a),   
is dominated by synchrotron emission (see also Sambruna 1997), as the 
X-ray spectra of the previously-observed HFSRQs in their database were
as steep as those of HBLs (Perlman et al. 1996b; Sambruna et al. 1996;
Padovani \& Giommi 1996).  As the DXRBS
sample contains a larger, more representative sample of HFSRQs than could be
gleaned from previously identified samples, we will revisit this assertion
and address the properties of the HFSRQ subclass in depth in a future paper
(Perlman \& Padovani, in prep). However, the data herein
allow us the first measure of the prevalence of such objects and their
proportion among FSRQs in a well defined sample, as well as the first
opportunity to speculate upon their relationship to the FSRQ subclass as a
whole.     

In order to examine the differences between the HFSRQs and lower 
$L_X/L_R$ objects, we have performed 1-dimensional K-S tests on the radio
and X-ray luminosity distributions on the subsamples of DXRBS FSRQs with
$L_X/L_R$ greater and less than $10^{-6}$.  These tests reveal that the
probability that the X-ray luminosity distribution of the two subsamples
could emerge from the same parent population is 41\% (i.e. consistent with
having been drawn from the same parent population), while the probability
that the radio luminosity distribution of the two subsamples
could emerge from the same parent population is $\sim 0.01\%$. The same story 
is
told by the mean X-ray and radio luminosities:  The mean radio luminosities
differ by 0.75 in the log and the significance of the difference is $>99.99\%$,
where as the difference in the X-ray luminosities is only 0.18, and is not
statistically significant ($P=77\%$).  This trend can also 
be seen on Figure 5. There is only one HFSRQ at  
luminosities $L_R > 10^{35} {\rm ~erg ~s^{-1} ~Hz^{-1}}$, compared to over 
two dozen lower-$L_X/L_R$ objects.  And a careful examination of the figure 
reveals that the lower-$L_X/L_R$ objects are much more strongly clustered
at high radio luminosities than are the HFSRQs.  The fraction of HFSRQs also
increases as radio luminosity decreases.  These trends are 
similar to (but not as marked) as what is seen for BL Lacs (Urry \& Padovani
1995; see also below).


\subsection {BL Lacs}

The BL Lac objects found within DXRBS fall mostly in the range $10^{-7.5} 
\lsim  L_X/L_R \lsim 10^{-5.5}$. Only a few objects are found at higher values
of $L_X/L_R$, as expected given our radio flux limits (\S~2.1).  A large 
fraction of the BL Lacs so far in our sample are at $L_X/L_R < 10^{-6.5}$,
the region of Figure 6 populated by LBLs.  In this region of parameter space
(to the left of the left-most dashed line in Figure 6), DXRBS includes 
objects up to two orders of magnitude fainter than the 1 Jy survey.
Most of these objects are radio galaxies (which are discussed in more detail
in \S 7), but three are clearly BL Lacs.
As with the FSRQs, this is expected given our much fainter flux limits.  This
comment can also be made for objects in the range  $10^{-6.5} \lsim L_X/L_R
\lsim 10^{-5.5}$, between the two dashed lines in Figure 6 -- of which all
but one object is classified herein as a BL Lac.

About 50\% of
the BL Lacs so far in our sample fall in the range $10^{-6.5} \lsim L_X/L_R
\lsim 10^{-5.5}$ ($0.72 \lsim \alpha_{\rm rx} \lsim 0.85$), and are 
``intermediate'' BL Lacs, objects with spectral shapes intermediate between
the X-ray bright and radio-bright varieties of BL Lacs (Padovani \& Giommi
1995a).  Similar objects have also been found in two ROSAT All-Sky
Survey-based samples (Kock et al. 1996, Nass et al. 1996), as well as the
{\it Einstein} Slew Survey sample (Perlman et al. 1996a).  For comparison,
we have plotted the $(\alpha_{\rm ox},\alpha_{\rm ro})$ values for members 
of these three ROSAT-based surveys (DXRBS, HQS/RASS and RC) in Figure 7.  
Since redshift information for the ROSAT-based samples is still incomplete,
it is difficult to compare them on the ($L_X,L_R$) plane.  It is important to
note that the optical magnitudes used to derive the effective
spectral indices at present include the galaxy contribution; this explains the
somewhat extreme objects in the lower right corner of the diagram, all BL Lacs
and radio galaxies 
at relatively low redshifts. If only the non-thermal flux were used,  
these objects would move towards the other objects along lines parallel to
the dashed lines in Fig. 7. 

As can be seen from Figure 7, each of these surveys covers a slightly different
region of the $(\alpha_{\rm ox},\alpha_{\rm ro})$ plane.  The HQS/RASS survey 
of  Nass et al. (1996) is dominated by objects at $\alpha_{\rm rx} < 0.72$ 
(the left-most diagonal line plotted in Figure 7), i.e. HBLs, but does
include a significant fraction of these intermediate objects (8 of 34).  Its
makeup is thus similar to the Slew Survey sample (Perlman et al. 1996a), which
contains 5 transition objects and 5 LBLs ($\alpha_{\rm rx} > 0.85$, the 
right-most line plotted in Figure 7) among a sample of 66 (Figure 6;
note that the diagonal lines thereon plotted represent the same values of
$\alpha_{\rm rx}$).  By comparison, objects from the RC survey of Kock et al. 
(1996) are more heavily concentrated (6 of 13) at 
intermediate values of $\alpha_{\rm rx}$,
although another 6 are HBLs.  The fraction of intermediate BL Lacs among the
DXRBS sample (15 of 32) is comparable to that in the RC sample.  However,
as shown in Figure 7, the DXRBS intermediate BL Lacs are concentrated at lower
values of $\alpha_{\rm ro}$ than those in either the HQS/RASS or RC samples. 
The large majority of the remainder (12 of 32) of the DXRBS BL Lacs are LBLs, 
and only a few objects (5 of 32) are HBLs.  

These differences are no doubt due
to the differing flux limits of the surveys.  The similarities 
of the HQS/RASS and Slew Survey samples are no surprise given their low radio
flux limits (a few mJy) and coverage of mostly X-ray bright objects.  By
comparison, the RC sample covers a range of X-ray fluxes similar to the 
HQS/RASS sample, but does not go as deep as DXRBS (by a factor $\sim 10$),
while its radio flux limit, at $\sim 40$ mJy, is similar to ours.  Finally,
both the HQS/RASS and RC survey groups observed only objects with optical
counterparts on sky survey plates, a restriction not found in DXRBS.  An 
examination of Figure 7 reveals that these facts naturally translate to the
$(\alpha_{\rm ox},\alpha_{\rm ro})$ plane.

What is most important in Figures 6 and 7 is that once again
the advantages of newer, deeper surveys which cover large dynamic ranges
of fluxes in more than one survey band is shown.  Only these very recent 
surveys (and particularly DXRBS, which already contains more intermediate 
BL Lacs than the HQS/RASS and RC samples combined) have revealed a large 
population of BL Lacs with $0.72 < \alpha_{\rm rx} < 0.85$; they went largely 
undetected in the 1 Jy and EMSS surveys because of the single-band
nature and small dynamic range covered by those surveys (Stickel et al. 1994,
Stocke et al. 1991, 1997).  
The exact population fraction of these ``intermediate'' BL Lacs
is not yet known, as a bivariate luminosity function has yet to be computed 
for the BL Lac class.  
Our results do not allow us to comment significantly on 
the relative proportion of HBLs and LBLs among BL
Lacs (e.g., Padovani \& Giommi 1995a; Urry \& Padovani 1995), since we are
sensitive to high $L_X/L_R$ objects only at high X-ray fluxes. 

\section {Selection Effects}

Several effects may bias samples of blazars, causing them to miss
objects which fall within their survey area and flux limits.   Many, but
not all, are intimately tied up with the question of classification (\S~3).
We will attempt to discuss each of these effects in turn.  They include the
the Browne \& 
March\~a (1993; BM) effect, lack of consistent (or consistently applied)
identification criteria in some samples,  and the effect of continuum 
variability on spectral indices.   Tied up with the second topic is the
question of whether all broad-line, flat-spectrum objects should be 
classified as FSRQs, or whether objects in which the continua are dominated by 
galactic light should be referred to as BLRG.

It is safe to say that there is probably no survey which is completely
immune from selection  effects.  The impact of selection effects upon
most previous surveys is not known, but must be understood to make progress
towards better understanding the AGN phenomenon.  The BM
(1993) effect has caused Perlman et al. (1996b) and Stocke et al.
(1997) to reconsider the makeup of the original Morris et al. (1991)
complete sample of EMSS BL Lacs, adding three BL Lacs to the C-EMSS
sample after an exhaustive perusal of the available X-ray, radio and
optical data, followed by further ROSAT HRI observations and optical 
spectroscopy.  These two papers represent the only serious attempts to
reformulate existing samples of blazars to minimize selection effects.
Other existing samples should be re-considered in similar fashion.

\subsection {The Browne \& March\~a Effect}

The BM effect
causes low-luminosity blazars (particularly BL Lacs) 
to be missed in surveys because the
apparent luminosity of the non-thermal nuclear source does not exceed
that of the host galaxy by a large factor.   X-ray and radio-faint surveys
are the most heavily affected.  The effect probably is more
severe for LBLs than HBLs, since galaxies hosting a relatively weak non-thermal
source of the LBL type (where the peak of the synchrotron emission is
at energies lower than 4000 \AA) might not qualify as a BL Lac simply because
its nonthermal emission at 4000 \AA~ is already much reduced compared to its
peak in the IR, and not strong enough to produce a Ca II H \& K break less
than 25\%.  Yet an HBL-type object with identical radio characteristics 
(flux, spectral shape, polarization, etc.) would be much more likely to be
classified as a BL Lac since its synchrotron emission at 4000 \AA~ is much
stronger and still growing.

We may gauge the impact of the BM effect upon our sample by
considering the likely properties of such objects (low-luminosity
blazars hidden within bright galaxies). In such a case, the optical
spectrum would resemble that of a radio galaxy, either with broad,
narrow, or no emission lines whatsoever  (a similar argument, although for 
BL Lacs only, was given in Laurent-Muehleisen et al. 1997).  We have attempted
to eliminate this ambiguity for broad-lined sources by grouping all broad-line
radio galaxies with the FSRQs (but see \S 7.2 below).  It therefore
remains for us to consider the narrow-lined objects which remain in our 
sample.

Three of our
first 85 IDs (WGAJ0340.8$-$1814, WGAJ0500.1$-$3040, WGAJ2317.9$-$4213)
meet this description 
and have herein been described as radio galaxies.
A few other objects which we have tentatively identified as BL Lacs (Table 5)
may fall into this classification when higher signal-to-noise spectra are 
taken (these have been individually discussed in \S 4.2). 
Ten of the previously identified sources also fall into
this category as NLRGs.  
But given that these objects exhibit flat-spectrum
radio sources, some of these objects may house low-luminosity 
blazars. This is particularly true of the ten previously identified
sources, for which the NLRG classifications (taken from the literature) were
made by older standards (usually -- but not always -- the classical definition
mentioned in \S 3).  Further observations should
be made to further probe their nature.  

We have plotted these objects in Figures 6 and 7. 
Inspection of Figure 6 reveals that
these NLRGs are less luminous than BL Lacs on average, but not the
least luminous objects in our sample.  
This result is consistent with the predictions made by
Browne \& March\~a (1993) and March\~a \& Browne (1995).  Perhaps more
interesting is the fact that the large majority (10 of 13) of these NLRGs 
lie within the LBL region of the $(\alpha_{\rm ox},\alpha_{\rm ro})$ plot 
(Figure 7).
This verifies our suspicion (above) that, since the synchrotron continuum 
produced by LBLs most often peaks in the near-to-mid infrared, and is already 
decreasing in the optical, the BM effect is stronger among LBLs than HBLs.  
Also noteworthy is the fact that the majority of these radio galaxies -- 7 of
13 -- are at low values of $\alpha_{\rm ro}$ and high values of 
$\alpha_{\rm ox}$.  As with the BL Lacs which occupy this region of Figure 7,
these objects are all at low $z$ and therefore the optical fluxes are largely
contaminated by the host galaxy.  

In addition to these objects, a large fraction of the newly-identified 
BL Lacs in our sample (10 of 22; see Table 5) have been so classified only by 
virtue of 
our usage of the expanded March\~a et al. (1996) definition of the BL Lac 
class.  These objects would not have been classified as BL Lacs under earlier,
more restrictive definitions of the class (Stickel et al. 1991, Stocke et al. 
1991, Perlman et al. 1996a), though they might have received some mention 
under such standards.  The large fraction of objects falling in this category
confirms the predictions of March\~a \& Browne (1995) for low-flux-limit X-ray
surveys such as this.  Similar results were also found in the RGB BL Lac 
Survey (Laurent-Muehleisen et al. 1997), as well as the 200 mJy sample 
(March\~a et al. 1996). Note that March\~a et al. showed that at least some 
of the objects outside
the ``classical'' BL Lac region of the $(C, W_\lambda)$ (but within the 
expanded March\~a et al. definition) also share the polarization properties of
BL Lacs.

Returning to Figure 7, it is now important to point out that we suspect that 
the two ROSAT-based samples to which we compared the DXRBS BL Lacs in \S 6.2,
those of Kock et al. (1996) and Nass et al. (1996), may contain a number
($\sim$ 20\%, as in the EMSS; Stocke et al. 1997) of objects which could be 
classified as BL Lacs
using the March\~a et al. (1996) redefinition of the BL Lac class.  This is 
because both Kock et al. and Nass et al. used (somewhat unclearly defined)
versions of the classical BL Lac-radio galaxy division to define their samples.
It is also possible that these two samples may contain a few radio galaxies
whose spectra should be more carefully scrutinized, as the regions of parameter
space that they cover overlap significantly with the radio galaxies in our
sample (Figure 7).  Further evidence for this point can be seen in the 
recent findings of Laurent-Muehleisen et al. (1997, specifically their Fig. 3).

\subsection{Inconsistent (or Inconsistently Applied) Identification Criteria}

As we mentioned in \S~3, there has not, to date, been either a single 
definition of the blazar class or of the BL Lac and FSRQ subclasses.  
The literature contains several examples of such inconsistencies.
For example, as pointed out by Perlman et al. (1996b), the lack of a Ca II
H \& K break strength criterion in the 1 Jy, S4 and S5 samples of BL
Lacs, has caused numerous objects to be misclassified as radio galaxies
instead of BL Lacs (see also March\~a \& Browne 1995, 1996).  It is therefore
likely that a number of low-$z$ objects may have been missed in this fashion.

Similarly,
inspection of the spectra of 1 Jy BL Lacs (Stickel et al. 1993) reveals
several which have emission lines considerably stronger than $W_\lambda = 
5(1+z)$ \AA.  While one may argue that some of these objects fall in the 
expanded region of the (H \& K break strength, equivalent width) diagram that 
March\~a et al. (1996) allot to the BL Lacs, several, for example 1 Jy 
1308+326, do not.  The majority of these broad-line 
objects are at $z>0.5$ and they may be the reason why the redshift distribution
of the 1 Jy BL Lac sample is so dissimilar from the EMSS, Slew and DXRBS
BL Lac redshift distribution.  The variable nature of blazars makes this
a particularly thorny problem to deal with.  It is well known that several
BL Lacs exhibit emission lines in their faint states (as we mentioned in 
\S~3).  Therefore it is entirely possible that the classification of a 
given object may be related to the state it was in when its classification
spectrum was taken.  

A third facet of this problem is the question of whether all flat-spectrum,
broad-lined objects deserve to be called FSRQs.  We believe this to be the
case, based upon the general similarities of properties between BLRGs and
FSRQs (e.g., Siebert et al. 1996). However we should note that the intrinsic 
power of the 
AGN affects how a source will be classified (see above).  Others, however,
have taken a different approach, and as a result some of the 
previously-identified objects which we list as FSRQs in Table 6 have been
classified as BLRGs by other authors. 

\subsection{The Effect of Continuum Variability on Spectral Indices}

Another possible bias, present in this as well as all other surveys which
use spectral indices based upon nonsimultaneous data in their selection
process, is due to variability.  Existing X-ray surveys have 
made variability-based allowances for this effect, which
likely decreases its impact greatly (see, e.g., Laurent-Muehleisen 1996;
Perlman et al. 1996a).  The magnitude of this effect upon existing
samples which used radio-spectral index criteria has been addressed by     
Drinkwater et al. (1997), who utilize the variability statistics compiled
by Stannard \& Bentley (1977) to estimate that the number of sources which have
average values of $\alpha_{\rm r} < 0.5$ that would not be included in a 
survey based upon two radio fluxes measured at different frequencies at times 
separated by $\sim 2$ years, is $\sim 10\%$. 

We have chosen to take a somewhat different approach to addressing this
issue.  The basis for this approach is not only that variability affects the
measured spectral index when the data points in question are non-simultaneous,
but also that the physical meaningfulness of a cut at $\alpha_{\rm r} = 0.5$, 
as opposed to, say, $\alpha_{\rm r} = 0.7$ has never truly been tested.
A factor of two variability between nonsimultaneous observations 
at 6 and 20 cm will change the observed radio spectral index by 0.58.  
In order to minimize the effect of non-simultaneous radio survey data
upon our samples, we decided to expand the common definition of
flat-spectrum radio sources to extend to $\alpha_{\rm r} = 0.7$ (instead of 
0.5).  Selecting all sources with $\alpha_{\rm r}
\leq 1.1$ (that is, 0.5 plus 0.6 to include a factor of two variability) 
would considerably lower the efficiency of the technique, as the
large majority of such steep-spectrum sources are radio galaxies
(although some are radio-loud quasars, often called SSRQs, or
steep-spectrum radio-loud quasars, which are thought to be oriented at
intermediate angles between FR 2s and FSRQs). The compromise approach
we adopted
allows an intrinsically $\alpha_{\rm r} = 0.5$ source to vary by $\sim 20\%$
between 20 cm and 6 cm survey observations.  We believe the
incompleteness due to missing sources which varied by larger amounts is
small ($<5\%$) given the distribution of instantaneous 3.6-6 cm spectral
indices  among core-dominated radio sources from our ATCA radio
survey, and we will use the ATCA data to both estimate the contamination from
truly steep-spectrum sources (and try to eliminate it) and test the 
significance and meaning of both our cutoff and the more traditional
$\alpha_{\rm r} = 0.5$ one.  

It is important to note that the selection of sources with $\alpha_{\rm
r}$ as high as 0.7 makes our BL Lac sample virtually 100\% complete. In fact,
of the 180 confirmed BL Lacs with radio spectral index information in the
multifrequency AGN database of Padovani et al. (1997b), only 5\% have 
$\alpha_{\rm r} > 0.7$ and all of these have X-ray-to-radio flux ratios much 
higher than those to which we are sensitive to. In other words, no BL Lac 
object should have been missed because of the $\alpha_{\rm r}$ cut. 

\section{Conclusions}

While the DXRBS sample is not yet completely identified (the objects discussed
in this paper represent $\sim 60\%$ of our object list), this paper has 
detailed a number of interesting and exciting results from our deep survey.
Most prominent among these results are:

1. A very high efficiency (95\%) at finding FSRQs and BL Lacs once the list
of radio-X-ray sources found by a cross-correlation of the ROSAT WGACAT with
single-dish radio catalogs has been limited to serendipitous flat radio 
spectrum sources ($\alpha_{\rm r} \leq 0.70$).

2. The DXRBS sample has vastly expanded coverage of the low luminosity end of 
the luminosity function both for BL Lacs and FSRQs, compared to all previous
samples of blazars.  Twenty-eight of 135 DXRBS FSRQs are at
$L_R < 10^{33.5} {\rm ~erg ~s^{-1} ~Hz^{-1}}$, compared to only 12 of 383
in the 1 Jy and S4 surveys combined.  Among these
28 DXRBS objects, six are at $L_R < 10^{32.5}{\rm ~erg ~s^{-1} ~Hz^{-1}}$. 
These numbers are 
sure to increase as the remaining 40\% of DXRBS objects (primarily
optically faint) are identified. 

For the BL Lacs, the increase is just as drastic, though restricted to 
objects with $L_X/L_R < 10^{-5.5}$ (i.e. LBL and intermediate BL Lacs).
The DXRBS sample includes eight BL Lacs with $L_R < 10^{32} 
{\rm ~erg ~s^{-1} ~Hz^{-1}}$ and $L_X/L_R < 10^{-5.5}$. While a few such 
objects are probably also included in the ROSAT based samples of Nass et al.
(1996) and Kock et al. (1996), their prevalence in these samples is 
difficult to evaluate because of the large fraction of objects in those 
samples which lack redshifts.  However it must be smaller given the
higher X-ray flux limits of the Kock et al. and Nass et al. surveys,
which are an order of magnitude higher than DXRBS.  This is confirmed
by the more recent work of Bade et al. (1997), who have just published 
redshifts for all but a few of the Nass et al. sample; they find very
few objects in their sample at $L_X < 10^{26}{\rm ~erg ~s^{-1} ~Hz^{-1}}$.  

3. DXRBS has also filled large holes in our coverage of $(L_X,L_R)$
parameter space, both for BL Lacs and FSRQs.  The impact here is much more
drastic for the FSRQs.  Prior to DXRBS, only nine FSRQs within complete 
samples were known at values of $L_X/L_R > 10^{-6}$.  Indeed, the continuity
of LBL and FSRQ broad-band and X-ray spectral properties led Sambruna et al.
(1996) to predict that no class of HBL-like FSRQs exists.  Our results clearly
refute this prediction.  Thirty-two of the 135 (25\%) DXRBS FSRQs so far
identified fall in this category; the fraction is even larger (40\%; 25 of 59)
among the newly identified objects.  These objects (whose numbers will surely
increase as the remainder of the DXRBS sample is identified), which we term 
HFSRQs, exhibit clearly smaller (by nearly an order of magnitude) radio 
luminosities 
than lower $L_X/L_R$ FSRQs.  In the light of this finding, a re-examination
of the broadband properties of FSRQs and indeed of the blazar class is in 
order.  We intend to make this subject a priority in our future work.

For BL Lacs, DXRBS contains a large number of ``intermediate'' BL Lacs, 
objects with $10^{-6.5}<L_X/L_R<10^{-5.5}$.  Until very recently, this 
region of parameter space was almost completely unexplored.  The {\it Einstein}
Slew survey found the first such objects (Perlman et al. 1996a), and more
recently two ROSAT based surveys (Kock et al. 1996, Nass et al. 1996) have
found considerable numbers of such objects.  However, due to the considerably
fainter flux limit of DXRBS, our sample includes fainter objects in this
region of parameter space than any previous sample.

\acknowledgements

EP acknowledges support from a USRA Visiting Scientist Fellowship while at
Goddard Space Flight Center, and helpful discussions with G. Madejski and 
C. M.  Urry.  PP acknowledges financial support from MURST and ASI. 
PP and PG acknowledge S. Benetti and M. Turatto for their assistance 
at La Silla, and M. Della Valle and A. Fontana for their help in preparing for
the ESO observing run. RS and LJ acknowledge the support of 
NRC Regular and Senior Research Fellowships while at Goddard Space Flight
Center. This work would have not been possible without the
availability of the many radio, 
optical, and X-ray databases quoted in this paper, namely the NVSS, PMN, GB6, 
NORTH20CM, TEXAS, APM, COSMOS, WGACAT. We thank all the persons involved in the
making of these databases for their effort. This research has made use of the
BROWSE program developed by the ESA/{\it EXOSAT} Observatory and by
NASA/HEASARC and of the NASA/IPAC Extragalactic Database (NED), which is
operated by the Jet Propulsion Laboratory, California Institute of Technology,
under contract with the National Aeronautics and Space Administration.   The
Australia Telescope Compact Array, a facility of the Australia Telescope
National Facility, is funded by the Commonwealth of Australia for operation
as a National Facility managed by CSIRO.


\newpage 
\centerline{\bf Figure Captions}


{\bf Figure 1.} Ca II break strengths $C$ and rest-frame 
emission line equivalent widths
of radio galaxies and BL Lacs are shown.  Quasars are not graphed here
because they fall too far to the right to be included (as does one radio 
galaxy, WGAJ0500.1$-$3040, which, despite the extremely large equivalent width
of several of its emission lines, is a narrow-line object, as described in \S
4).  We have overplotted the traditional definition of the BL Lac class (dashed
box) as used in Stickel et al. (1991), Stocke et al. (1991) and Perlman et al.
(1996a), as well as the expanded definition of the BL Lac class advocated by
March\~a et al. (1996) (the region in between the dot-dashed lines).  Objects
where both $C$ and $W_\lambda$ could be measured are shown as squares.  
Objects where one or both of these figures are
upper limits are denoted by diamonds.  All error bars shown are 1 $\sigma$, and
all upper limits shown are at the 2 $\sigma $ significance level.

{\bf Figure 2.} Optical spectra of all 85 objects for which we announce 
identifications in this paper.  All spectra have been dereddened and cleaned
of cosmic rays as described in Section 3.

{\bf Figure 3.} Redshift distribution for the DXRBS, S4, and 1 Jy samples of
FSRQs (radio quasars with $\alpha_{\rm r} \le 0.7$).  

{\bf Figure 4.}  Redshift distribution for the DXRBS, Slew, and 1 Jy samples
of BL Lacs. The hatched areas represent lower limits.    Redshift figures for
1 Jy BL Lacs have been taken from Stickel et al. (1994), while those for Slew
BL Lacs have been taken from Perlman et al. (1996a),  Bade et al. (1997), and
Perlman, Schachter \& Stocke (in preparation).

{\bf Figure 5.} The X-ray and radio luminosities of FSRQs. Newly 
identified DXRBS FSRQs are shown as filled circles, while previously identified
serendipitous DXRBS FSRQs are shown as filled squares. DXRBS objects identified
as radio galaxies with broad emission lines are shown as crosses. 
The published complete samples of blazars (the 1 Jy [triangles] and S4 
[squares]) cover the 
low-luminosity end very poorly: while still incomplete, 
the DXRBS blazar survey already includes higher numbers of faint
FSRQs (over 3$\times$ the number in the 1 Jy and S4 combined).  One in 
4 FSRQs have high ratios of X-ray to radio luminosity $L_X/L_R > 10^{-6}$ 
(to the right of the dashed line).  Previous radio surveys included very
few objects in this region.
See sections 5 
and 6 for discussion. Radio data for the S4 and 1 Jy sources from Stickel \& 
K\"uhr (1994) and Stickel et 
al. (1994); X-ray data from the multifrequency AGN database of Padovani et al. 
(1997b) and references therein. Note that X-ray data are 
available only for $\sim 53\%$ and $66\%$ of the S4 and 1 Jy FSRQs
respectively.

{\bf Figure 6.} The X-ray and radio luminosities of BL Lacs. Newly 
identified DXRBS BL Lacs are shown as filled circles, while previously 
identified serendipitous DXRBS BL Lacs are shown as filled squares. 
DXRBS objects identified as radio galaxies with narrow or no emission 
lines are shown as crosses. The 1 Jy, Slew, and EMSS BL Lacs are represented
by triangles, circles, and squares respectively. Crosses represent the
NLRGs in our sample.
While DXRBS does not include extremely 
high $L_x/L_r$ BL Lacs such as those found in the {\it Einstein} Slew 
Survey, it can be seen that prior to DXRBS, region of the graph between 
$-6.5 \lsim \log L_x / L_r \lsim -5.5$ (denoted by two dashed lines) 
was very poorly populated, a consequence of the disparate survey methods
used.  The high sensitivity and combined selection method of DXRBS 
reveals the previous ``zone of avoidance'' in this graph to be illusory.
See Sections 4 and 5 for discussion. Most of the data come from the original
papers; additional radio and X-ray data are from the multifrequency AGN 
database of Padovani et al. (1997) and references therein. 

{\bf Figure 7.} The X-ray-optical ($\alpha_{ox}$) and radio-optical
($\alpha_{ro}$) effective spectral indices of the BL Lacs in the DXRBS sample
compared to those in the samples of Kock et al. (1996) and Nass et al. (1996).
Newly identified DXRBS BL Lacs are shown as filled circles, while previously
identified serendipitous DXRBS BL Lacs are shown as filled squares. Empty
circles represent the Nass et al.'s objects, while empty squares indicate 
the Koch
et al.'s sources. Crosses represent the
NLRGs in our sample. The two dashed lines denote the loci of points with $\log
L_x / L_r = -6.5$ ($\alpha_{\rm rx} \simeq 0.85$, upper line) and $\log L_x /
L_r = -5.5$ ($\alpha_{\rm rx} \simeq 0.72$, lower line). Each of the three
surveys covers different areas of parameter space, as shown. The spectral
indices $\alpha_{ox}$ and $\alpha_{ro}$ are defined in the usual way and
calculated between the rest-frame frequencies of 5 GHz, 5000 \AA, and 1 keV.

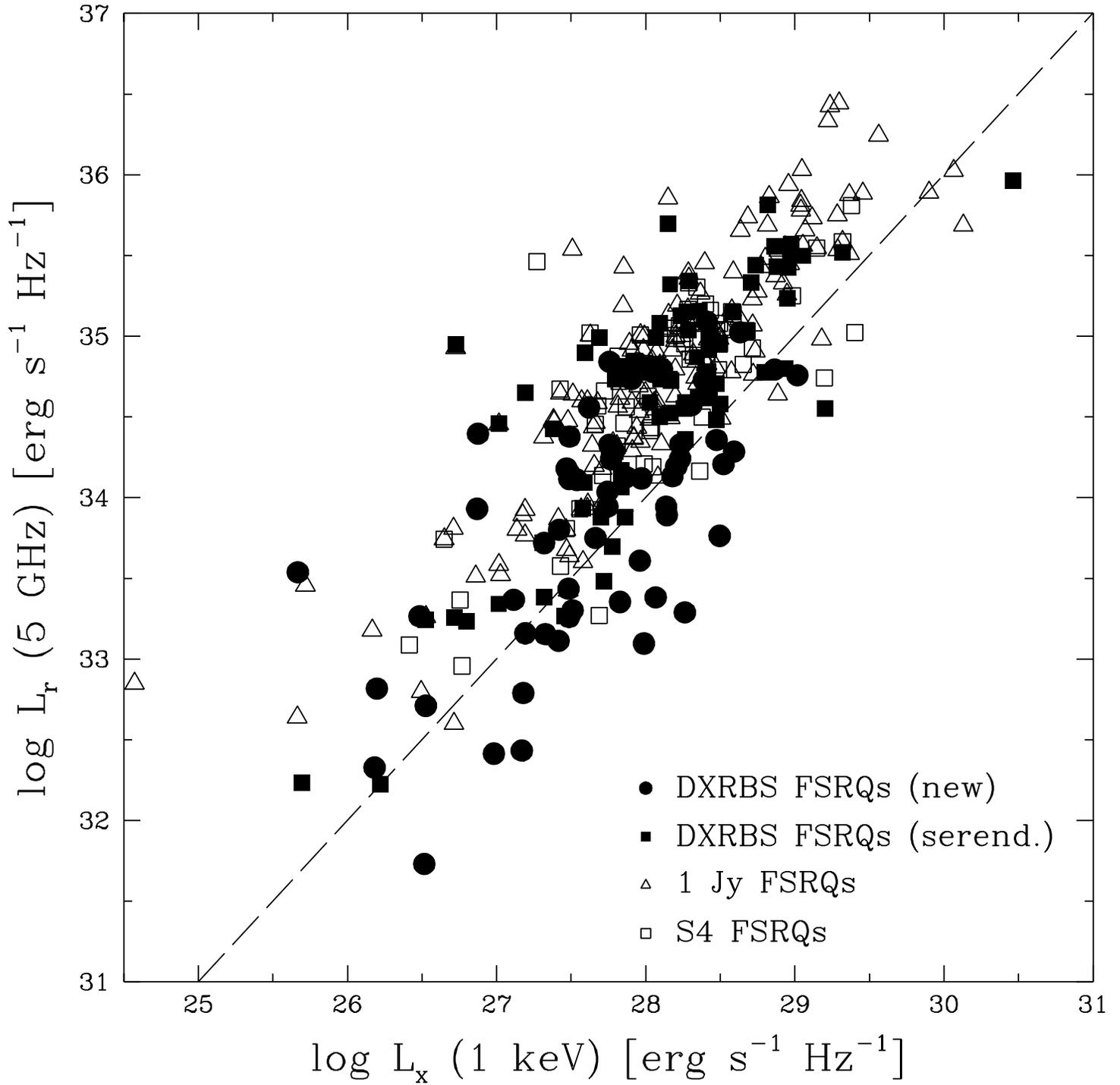
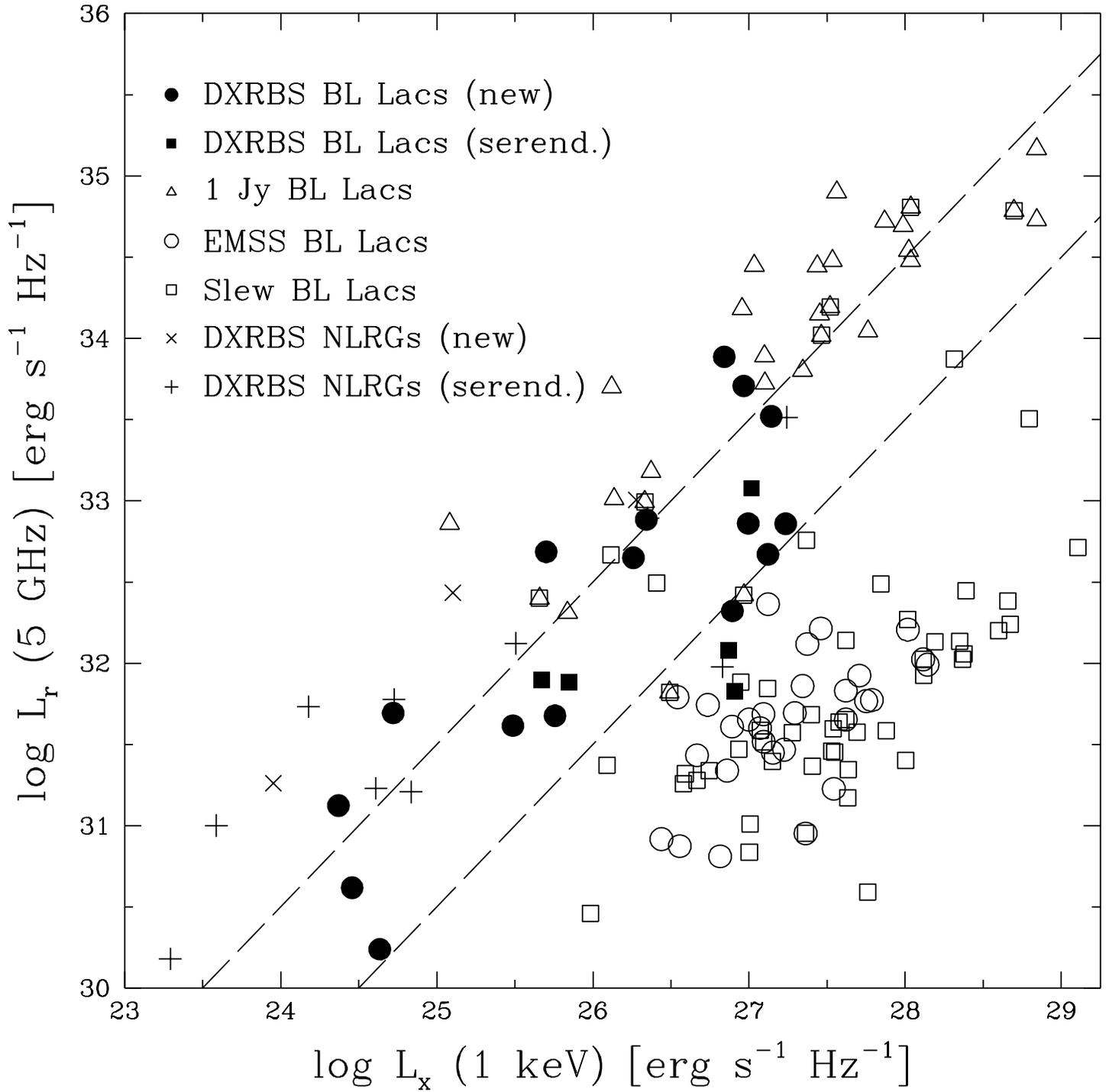
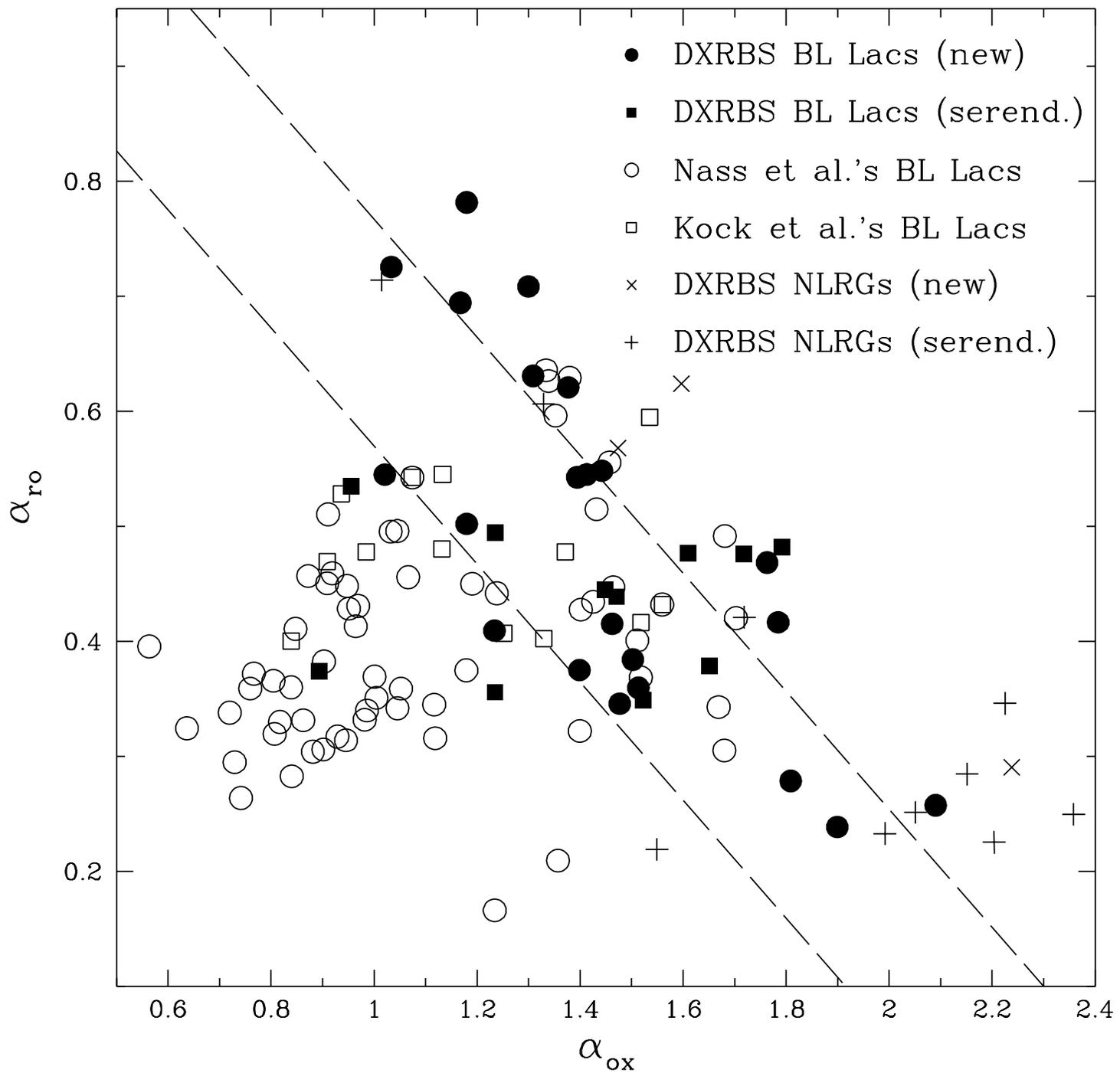
\end{document}


\begin{deluxetable}{lccl}
\tablecaption{Log of Observations}
\tablewidth{0pc}
\tablecolumns{4}
\tablehead{
\colhead{Name} & \colhead{Where} & \colhead {Date} & \colhead{Observers} \\
& \colhead{Observed} & \colhead{(MM/YY)} }
\startdata

WGAJ0011.2$-$3620 & ES 3.6m & 12/96 & Giommi, Padovani \nl
WGAJ0012.5$-$1629 & KP 2.1m & 12/96 & Perlman, Sambruna \nl
WGAJ0015.5+3052 & MMT     & 12/96 & Perlman \nl
WGAJ0029.0+0509 & KP 2.1m & 12/96 & Perlman, Sambruna \nl
WGAJ0032.5$-$2849 & ES 3.6m & 12/96 & Giommi, Padovani \nl
WGAJ0032.5$-$2648 & KP 2.1m & 12/96 & Perlman, Sambruna \nl
WGAJ0043.3$-$2638 & CT 1.5m & 01/97 & Perlman \nl
WGAJ0049.5$-$2509 & ES 3.6m & 12/96 & Giommi, Padovani \nl
WGAJ0057.3$-$2212 & ES 3.6m & 12/96 & Giommi, Padovani \nl
WGAJ0100.1$-$3337 & ES 3.6m & 12/96 & Giommi, Padovani \nl
WGAJ0110.5$-$1647 & KP 2.1m & 12/96 & Perlman, Sambruna \nl
WGAJ0125.0+0146 & KP 2.1m & 12/96 & Perlman, Sambruna \nl
WGAJ0136.0$-$4044 & ES 3.6m & 12/96 & Giommi, Padovani \nl
WGAJ0143.2$-$6813 & ES 3.6m & 12/96 & Giommi, Padovani \nl
WGAJ0204.8+1514 & ES 3.6m & 12/96 & Giommi, Padovani \nl
WGAJ0210.0$-$1004 & ES 2.2m & 12/96 & Giommi, Padovani \nl
WGAJ0216.6$-$7331 & ES 3.6m & 12/96 & Giommi, Padovani \nl
WGAJ0217.7$-$7347 & ES 3.6m & 12/96 & Giommi, Padovani \nl
WGAJ0245.2+1047 & MMT     & 12/96 & Perlman \nl
WGAJ0304.9+0002 & MMT     & 12/96 & Perlman \nl
WGAJ0313.9+4115 & KP 2.1m & 12/96 & Perlman, Sambruna \nl
WGAJ0314.4$-$6548 & CT 1.5m & 01/97 & Perlman \nl
WGAJ0322.2$-$5042 & ES 2.2m & 12/96 & Giommi, Padovani \nl
WGAJ0324.9$-$2140 & ES 2.2m & 12/96 & Giommi, Padovani \nl
WGAJ0325.0$-$4926 & ES 2.2m & 12/96 & Giommi, Padovani \nl
WGAJ0340.8$-$1814 & ES 3.6m & 12/96 & Giommi, Padovani    \nl
WGAJ0357.6$-$4158 & ES 3.6m & 12/96 & Giommi, Padovani \nl
WGAJ0411.0$-$1637 & KP 2.1m & 12/96 & Perlman, Sambruna \nl
WGAJ0414.0$-$1307 & ES 3.6m & 12/96 & Giommi, Padovani \nl
WGAJ0414.0$-$1224 & ES 2.2m & 12/96 & Giommi, Padovani \nl
WGAJ0421.5+1433 & ES 2.2m & 12/96 & Giommi, Padovani \nl
		& CT 1.5m & 01/97 & Perlman \nl
WGAJ0428.8$-$3805 & ES 2.2m & 12/96 & Giommi, Padovani \nl
WGAJ0434.3$-$1443 & ES 3.6m & 12/96 & Giommi, Padovani \nl
WGAJ0435.1$-$0811 & KP 2.1m & 12/96 & Perlman, Sambruna \nl
WGAJ0447.9$-$0322 & KP 2.1m & 12/96 & Perlman, Sambruna \nl
WGAJ0448.2$-$2110 & KP 2.1m & 12/96 & Perlman, Sambruna \nl
WGAJ0449.4$-$4349 & CT 4m   & 01/96 & Ruiz(1) \nl
WGAJ0500.0$-$3040 & ES 2.2m & 12/96 & Giommi, Padovani \nl
WGAJ0502.5+1338 & ES 3.6m & 12/96 & Giommi, Padovani \nl
WGAJ0510.0+1800 & ES 3.6m & 12/96 & Giommi, Padovani \nl
WGAJ0513.8+0156 & ES 2.2m & 12/96 & Giommi, Padovani \nl
WGAJ0518.2+0624 & ES 3.6m & 12/96 & Giommi, Padovani \nl
WGAJ0528.5$-$5820 & CT 1.5m & 01/97 & Perlman \nl
WGAJ0535.1$-$0239 & ES 2.2m & 12/96 & Giommi, Padovani \nl
WGAJ0539.0$-$3427 & ES 2.2m & 12/96 & Giommi, Padovani \nl
WGAJ0544.1$-$2241 & KP 2.1m & 12/96 & Perlman, Sambruna \nl
		& ES 2.2m & 12/96 & Giommi, Padovani \nl
WGAJ0546.6$-$6415 & CT 1.5m & 01/97 & Perlman \nl
WGAJ0558.1+5328 & KP 2.1m & 12/96 & Perlman, Sambruna \nl
WGAJ0600.5$-$3937 & ES 2.2m & 12/96 & Giommi, Padovani \nl
WGAJ0624.7$-$3230 & CT 1.5m & 01/97 & Perlman \nl
WGAJ0628.4$-$3208 & ES 3.6m & 12/96 & Giommi, Padovani \nl
WGAJ0631.9$-$5404 & CT 1.5m & 01/97 & Perlman	 \nl
WGAJ0633.1$-$2333 & ES 3.6m & 12/96 & Giommi, Padovani \nl
WGAJ0648.2$-$4347 & ES 2.2m & 12/96 & Giommi, Padovani \nl
WGAJ0656.3$-$2403 & CT 1.5m & 01/97 & Perlman \nl
WGAJ0724.3$-$0715 & CT 1.5m & 01/97 & Perlman \nl
WGAJ0744.8+2920 & CT 1.5m & 01/97 & Perlman \nl
WGAJ0748.2$-$5257 & ES 2.2m & 12/96 & Giommi, Padovani \nl
WGAJ0750.9$-$6726 & ES 2.2m & 12/96 & Giommi, Padovani \nl
WGAJ0816.0$-$0736 & ES 2.2m & 12/96 & Giommi, Padovani \nl
                & CT 1.5m & 01/97 & Perlman \nl
WGAJ0900.2$-$2817 & ES 3.6m & 12/96 & Giommi, Padovani \nl
WGAJ0940.2+2603 & KP 2.1m & 12/96 & Perlman, Sambruna \nl
WGAJ1003.9+3244 & KP 2.1m & 12/96 & Perlman, Sambruna \nl
WGAJ1011.5$-$0423 & ES 2.2m & 12/96 & Giommi, Padovani \nl
WGAJ1025.9+1253 & KP 2.1m & 12/96 & Perlman, Sambruna \nl
WGAJ1032.1$-$1400 & CT 1.5m & 01/97 & Perlman \nl
WGAJ1035.0+5652 & KP 2.1m & 12/96 & Perlman, Sambruna \nl
WGAJ1046.3+5354 & KP 2.1m & 12/96 & Perlman, Sambruna \nl
WGAJ1057.6$-$7724 & ES 3.6m & 12/96 & Giommi, Padovani \nl
WGAJ1104.8+6038 & KP 2.1m & 12/96 & Perlman, Sambruna \nl
WGAJ1108.1$-$7748 & ES 3.6m & 12/96 & Giommi, Padovani \nl
WGAJ1112.5$-$3745 & ES 2.2m & 12/96 & Giommi, Padovani \nl
WGAJ1150.4+0156 & ES 3.6m & 12/96 & Giommi, Padovani \nl
WGAJ1222.6+2934 & CT 1.5m & 01/97 & Perlman \nl
WGAJ1224.5+2613 & KP 2.1m & 12/96 & Perlman, Sambruna \nl
WGAJ1231.7+2848 & CT 1.5m & 01/97 & Perlman \nl
WGAJ1300.7$-$3253 & CT 1.5m & 01/97 & Perlman \nl
WGAJ1324.0$-$3623 & CT 1.5m & 01/97 & Perlman	 \nl	
WGAJ1525.3+4201 & Lick 3m & 06/95 & Jones \nl
WGAJ2258.3$-$5525 & ES 3.6m & 12/96 & Giommi, Padovani \nl
WGAJ2317.4$-$4213 & ES 3.6m & 12/96 & Giommi, Padovani \nl
WGAJ2322.0+2114 & KP 2.1m & 12/96 & Perlman, Sambruna \nl
WGAJ2333.2$-$0131 & ES 3.6m & 12/96 & Giommi, Padovani \nl
WGAJ2347.6+0852 & KP 2.1m & 12/96 & Perlman, Sambruna \nl
WGAJ2350.6+3622 & KP 2.1m & 12/96 & Perlman, Sambruna \nl

\enddata
\end{deluxetable}


\newdimen\digitwidth
\setbox0=\hbox{\rm0}
\digitwidth=\wd0
\catcode `@=\active
\def@{\kern\digitwidth}

\setcounter{page}{46}
\setcounter{table}{5}
\begin{deluxetable}{lccllrrlcl}
\tablecaption{Previously Identified Blazars}
\tablewidth{0pc}
\tablecolumns{10}
\tablehead{
&\colhead{ROSAT}&\colhead{F(0.1$-$2.0 keV)}&\colhead{F(1 keV)}&\colhead{log} &
\colhead{F(6 cm)} \\
\colhead{Name}&\colhead{ct/s}&\colhead{${\rm erg ~cm^{-2} ~s^{-1}}$}&
\colhead{$\mu$Jy}&\colhead{$F_{\rm X}/F_{\rm R}$}&\colhead{mJy}&\colhead{$\alpha_r$}&
\colhead{V}&\colhead{Class}&\colhead{$z$} }

\startdata

PKS 0027$-$426    &0.046&0.55E$-$12&@0.14  &$-$11.80&  419& 0.00  &19.0&FSRQ&1.660\nl
PKS 0035+23       &0.012&0.16E$-$12&@0.032 &$-$12.49&  463& 0.54  &19.0&FSRQ&2.270\nl
PKS 0100$-$270    &0.030&0.27E$-$12&@0.045 &$-$11.92&  201& 0.42  &18.1&FSRQ&1.597\nl
TEX 0109+200      &0.017&0.24E$-$12&@0.058 &$-$12.01&  281& 0.31  &17.0&FSRQ&0.746\nl
PKS 0112$-$017    &0.026&0.36E$-$12&@0.092 &$-$12.56& 1540&$-$0.48&17.4&FSRQ&1.381\nl
PKS 0119+041      &0.024&0.26E$-$12&@0.050 &$-$12.70& 1235&$-$0.24&19.5&FSRQ&0.637\nl
UM 320            &0.010&0.12E$-$12&@0.031 &$-$12.22&  259& 0.13  &18.6&FSRQ&2.280\nl
PKS 0122$-$003    &0.066&0.76E$-$12&@0.20  &$-$12.11& 1234& 0.09  &16.7&FSRQ&1.070\nl
I 0115            &0.005&0.65E$-$13&@0.019 &$-$12.82&  671& 0.64  &14.0&RG&0.043\nl
PKS 0142$-$278    &0.050&0.56E$-$12&@0.14  &$-$12.01&  833&$-$0.03&19.0&FSRQ&1.157\nl
S4 0206+355       &0.022&0.31E$-$12&@0.086 &$-$12.39&  976& 0.68  &13.0&RG&0.037\nl
PKS 0247$-$207    &0.029&0.33E$-$12&@0.092 &$-$11.90&  389& 0.54  &15.5&RG&0.087\nl
PKS 0256$-$005    &0.023&0.34E$-$12&@0.10  &$-$12.04&  500&$-$0.59&17.2&FSRQ&1.995\nl
PKS 0335$-$364    &0.048&0.49E$-$12&@0.070 &$-$12.19&  583&$-$0.39&18.0&FSRQ&1.537\nl
PKS 0406$-$127    &0.013&0.13E$-$12&@0.021 &$-$12.63&  517& 0.22  &19.0&FSRQ&1.563\nl
PKS 0422$-$380    &0.064&0.67E$-$12&@0.011 &$-$12.49& 1706&$-$2.06&18.1&FSRQ&0.782\nl
PKS 0439$-$433    &0.100&0.10E$-$11&@0.16  &$-$11.53&  285& 0.19  &16.4&FSRQ&0.593\nl
PKS 0514$-$459    &0.067&0.84E$-$12&@0.16  &$-$12.10&  990& 0.44  &17.5&FSRQ&0.194\nl
S4 0537+531       &0.007&0.83E$-$13&@0.040 &$-$12.82&  668&$-$0.02&18.0&FSRQ&1.275\nl
EXO0556.4$-$3838  &0.956&0.11E$-$10&@3.01  & $-$9.73&   68& 0.00  &17.1&BL Lac&?\nl
S5 0743+74        &0.009&0.12E$-$12&@0.034 &$-$12.45&  445&$-$0.19&19.3&FSRQ&1.629\nl
0822+27W01        &0.054&0.64E$-$12&@0.16  &$-$11.31&  150&$-$0.03&17.7&FSRQ&2.060\nl
B2 0834+25        &0.035&0.41E$-$12&@0.11  &$-$11.94&  458& 0.14  &18.0&FSRQ&1.122\nl
TEX 0836+182      &0.014&0.11E$-$12&@0.016 &$-$12.52&  310& 0.17  &17.0&BL Lac&?\nl
PKS 0839+187      &0.051&0.54E$-$12&@0.104 &$-$12.23&  898& 0.24  &16.4&FSRQ&1.272\nl
S4 0847+37        &0.043&0.64E$-$12&@0.15  &$-$11.75&  391& 0.37  &19.5&RG&0.407\nl
4C 58.17          &0.023&0.28E$-$12&@0.071 &$-$12.51& 1184& 0.15  &18.0&FSRQ&1.322\nl
OJ$-$297          &0.004&0.58E$-$13&@0.022 &$-$13.40& 2216&$-$0.17&16.6&FSRQ&2.160\nl
4C 47.29          &0.032&0.27E$-$12&@0.044 &$-$12.73& 1264& 0.48  &18.7&FSRQ&1.462\nl
RXJ09168+5238     &0.252&0.22E$-$11&@0.36  &$-$10.55&   69& 0.58  &18.9&BL Lac&0.190\nl
S5 0916+718       &0.026&0.36E$-$12&@0.094 &$-$11.78&  295& 0.22  &19.5&FSRQ&0.594\nl
4C 55.17          &0.076&0.46E$-$12&@0.034 &$-$12.94& 2015& 0.33  &16.5&FSRQ&0.909\nl
S4 0955+476       &0.036&0.34E$-$12&@0.028 &$-$12.77& 1005&$-$0.32&18.0&FSRQ&1.873\nl
0959+68W01        &0.029&0.42E$-$12&@0.11  &$-$11.18&   86& 0.41  &15.9&FSRQ&0.773\nl
WGAJ1012.2+063    &0.021&0.21E$-$12&@0.055 &$-$12.03&  300& 0.45  &16.8&BL Lac&?\nl
4C 19.34          &0.103&0.12E$-$11&@0.28  &$-$11.75&  765&$-$0.11&17.5&FSRQ&0.828\nl
S5 1027+74        &0.066&0.84E$-$12&@0.22  &$-$11.38&  250& 0.19  &17.2&FSRQ&0.123\nl
B2 1048+34        &0.011&0.86E$-$13&@0.015 &$-$12.56&  312& 0.44  &19.0&FSRQ&2.520\nl
5C 02.56          &0.004&0.32E$-$13&@0.006 &$-$12.52&  100& 0.50  &20.0&FSRQ&2.396\nl
4C 72.16          &0.034&0.49E$-$12&@0.13  &$-$12.03&  858& 0.44  &17.9&FSRQ&1.460\nl
1104+72W01        &0.008&0.99E$-$13&@0.026 &$-$12.26&  271& 0.30  &18.9&FSRQ&2.100\nl
PKS 1145$-$071    &0.031&0.37E$-$12&@0.093 &$-$12.27&  791& 0.00  &18.0&FSRQ&1.342\nl
B2 1147+245       &0.017&0.15E$-$12&@0.022 &$-$12.72&  645&$-$0.02&16.2&BL Lac&$>0.2$\nl
WGAJ1202.1+444    &0.210&0.13E$-$11&@0.11  &$-$10.98&   69& 1.43  &18.1&BL Lac&?\nl
7C 1159+2813      &0.043&0.34E$-$12&@0.036 &$-$11.70&  107& 0.16  & ...&FSRQ&0.672\nl
DW 1211+13        &0.011&0.98E$-$13&@0.023 &$-$12.89&  894& 0.42  &18.1&FSRQ&1.137\nl
B2 1211+33        &0.039&0.25E$-$12&@0.041 &$-$12.44&  627& 0.54  &17.9&FSRQ&1.598\nl
1ES1212+078       &0.506&0.54E$-$11&@0.95  &$-$10.37&  117&$-$0.04&16.0&BL Lac&0.136\nl
B2 1215+33        &0.007&0.49E$-$13&@0.004 &$-$12.67&  117& 0.46  &18.1&FSRQ&2.605\nl
ON 325            &0.382&0.27E$-$11&@0.27  &$-$11.44&  478& 0.09  &15.6&BL Lac&0.130\nl
3C 273            &9.060&0.91E$-$10&17.2   &$-$11.68&43572& 0.12  &12.8&FSRQ&0.158\nl
GB 1231+482       &0.140&0.86E$-$12&@0.036 &$-$11.93&  268& 0.27  &17.4&FSRQ&0.375\nl
RX J12368+2507    &0.083&0.81E$-$12&@0.10  &$-$11.26&  109& 0.54  &17.6&FSRQ&0.546\nl
PKS 1240$-$294    &0.042&0.59E$-$12&@0.17  &$-$11.83&  472& 0.31  &17.7&FSRQ&1.133\nl
PKS 1255$-$316    &0.019&0.23E$-$12&@0.061 &$-$12.74& 1410& 0.09  &18.7&FSRQ&1.924\nl
1308+328          &0.014&0.12E$-$12&@0.027 &$-$12.76&  870&$-$0.62&19.5&FSRQ&1.650\nl
PKS 1324$-$300    &0.031&0.40E$-$12&@0.097 &$-$11.96&  406& 0.59  &18.0&RG&0.200\nl
NGC 5232          &0.010&0.11E$-$12&@0.020 &$-$12.72&  529&$-$0.09&13.0&RG&0.021\nl
1339+274          &0.027&0.19E$-$12&@0.021 &$-$12.44&  347&$-$0.29&19.0&FSRQ&1.185\nl
PKS 1402+044      &0.012&0.13E$-$12&@0.031 &$-$12.80& 1008&$-$0.49&19.0&FSRQ&3.202\nl
S4 1435+638       &0.078&0.89E$-$12&@0.13  &$-$12.02&  757& 0.51  &16.1&FSRQ&2.068\nl
MS 14428+6344     &0.022&0.17E$-$12&@0.030 &$-$12.43&  442& 0.36  &17.2&FSRQ&1.380\nl
3C 309.1          &0.077&0.78E$-$12&@0.17  &$-$12.63& 3567& 0.64  &16.8&FSRQ&0.904\nl
PKS 1502+106      &0.023&0.22E$-$12&@0.032 &$-$13.14& 2325&$-$0.33&18.6&FSRQ&0.563\nl
B2 1506+33A       &0.007&0.55E$-$13&@0.005 &$-$12.57&  115& 0.14  &18.5&FSRQ&2.200\nl
MC 1524+101       &0.024&0.24E$-$12&@0.043 &$-$12.15&  316& 0.26  &19.0&FSRQ&1.358\nl
1555+3538         &0.226&0.28E$-$11&@0.55  &$-$10.44&   80& 0.68  &14.7&RG&0.158\nl
NGC6034		  &0.032&0.38E$-$12&@0.077 &$-$11.99&  343& 0.58  &13.5&RG&0.034 \nl
NGC6061           &0.048&0.52E$-$12&@0.10  &$-$11.71&  258& 0.61  &13.6&RG&0.038\nl
OS319             &0.103&0.12E$-$11&@0.22  &$-$12.31& 2324& 0.18  &17.5&FSRQ&1.401\nl
4C 38.41          &0.100&0.10E$-$11&@0.15  &$-$12.61& 3221&$-$0.44&18.0&FSRQ&1.814\nl
S4 1637+574       &0.105&0.10E$-$11&@0.19  &$-$12.24& 1750&$-$0.29&17.0&FSRQ&0.750\nl
S4 1638+398       &0.028&0.26E$-$12&@0.050 &$-$12.61& 1117&$-$0.44&16.5&FSRQ&1.666\nl
3C 345            &0.317&0.35E$-$11&@0.75  &$-$12.31& 8719&$-$0.08&16.0&FSRQ&0.594\nl
B2 1656+34        &0.030&0.26E$-$12&@0.046 &$-$12.28&  474& 0.02  &18.5&FSRQ&1.936\nl
S4 1656+571       &0.021&0.20E$-$12&@0.037 &$-$12.61&  764& 0.04  &17.6&FSRQ&1.293\nl
4C 45.34          &0.042&0.55E$-$12&@0.17  &$-$11.65&  461& 0.70  &17.4&FSRQ&0.646\nl
PKS 1725+044      &0.049&0.70E$-$12&@0.18  &$-$12.07&  885&$-$0.44&17.0&FSRQ&0.293\nl
7C 1753+6543      &0.020&0.23E$-$12&@0.046 &$-$11.92&  186& 0.61  &17.4&FSRQ&0.140\nl
S4 1806+456       &0.027&0.30E$-$12&@0.077 &$-$11.99&  351&$-$0.68&19.0&FSRQ&0.830\nl
EXO1811.7+3143    &0.036&0.48E$-$12&@0.090 &$-$11.45&  127& 0.17  &17.4&BL Lac&0.117\nl
4C 50.47          &0.023&0.38E$-$12&@0.11  &$-$11.93&  345& 0.53  &17.9&FSRQ&1.098\nl
PKS 1937$-$101    &0.029&0.46E$-$12&@0.18  &$-$12.15&  750& 0.09  &17.0&FSRQ&3.787\nl
PKS 2058$-$425    &0.097&0.10E$-$11&@0.20  &$-$11.86&  721& 0.56  &17.2&FSRQ&0.221\nl
MH 2136$-$428     &0.120&0.97E$-$12&@0.11  &$-$11.17&  108& 0.00  &16.2&BL Lac&?\nl
4C 06.69          &0.186&0.33E$-$11&@0.91  &$-$12.02& 4135&$-$0.30&16.5&FSRQ&0.990\nl
OY$-$106          &0.008&0.72E$-$13&@0.016 &$-$13.73& 4271& 0.36  &19.5&FSRQ&0.618\nl
PKS 2212$-$299    &0.032&0.21E$-$12&@0.022 &$-$12.54&  450& 0.30  &17.4&FSRQ&2.706\nl
3C 446            &0.088&0.12E$-$11&@0.32  &$-$12.62& 6382& 0.00  &18.0&FSRQ&1.404\nl
I 1459            &0.084&0.88E$-$12&@0.15  &$-$12.16& 1151&$-$0.14&11.8&RG&0.006\nl
PKS 2316$-$423    &0.207&0.17E$-$11&@0.34  &$-$11.52&  595& 0.43  &14.5&BL Lac&0.055\nl
PKS 2319+07       &0.013&0.19E$-$12&@0.047 &$-$12.63&  857& 0.18  &18.5&FSRQ&2.090\nl
PKS 2329$-$384    &0.019&0.19E$-$12&@0.048 &$-$12.47&  784&$-$0.05&17.0&FSRQ&1.195\nl
PG 2344+092       &0.085&0.13E$-$11&@0.36  &$-$11.97& 1392& 0.32  &16.0&FSRQ&0.673\nl
PKS 2352$-$342    &0.182&0.12E$-$11&@0.11  &$-$11.53&  241& 0.62  &16.4&FSRQ&0.702\nl
PKS 2357$-$318    &0.037&0.32E$-$12&@0.085 &$-$12.19&  747&$-$1.62&17.6&FSRQ&0.991\nl
PKS 2357$-$326    &0.007&0.41E$-$13&@0.003 &$-$13.37&  535& 0.02  &18.7&FSRQ&1.275\nl

\enddata
\end{deluxetable}